\documentclass[aps,prb,twocolumn,showpacs,superscriptaddress,floatfix,nofootinbib]{revtex4}
\usepackage{graphicx}
\usepackage{amsmath}
\usepackage{epsfig}
\usepackage{helvet} 
\usepackage{amssymb}

\newcommand{\be}{\begin{equation}}
\newcommand{\ee}{\end{equation}}
\newcommand{\bea}{\begin{eqnarray}}
\newcommand{\eea}{\end{eqnarray}}

\newcommand{\tr}{\mbox{tr}}
\newcommand{\bra}[1]{\mbox{$\langle #1 |$}}
\newcommand{\expect}[1]{\mbox{$\langle #1 \rangle$}}

\newcommand{\ket}[1]{\mbox{$| #1 \rangle$}}

\newcommand{\hc}{\hat c}
\newcommand{\hcdag}{\hat c^\dagger}
\newcommand{\hw}{\hat w}
\newcommand{\hu}{\hat u}
\newcommand{\hwdag}{\hat w^\dagger}
\newcommand{\hudag}{\hat u^\dagger}

\newcommand{\hP}{\hat P}
\newcommand{\hH}{\hat H}

\newcommand{\hO}{{\hat O}}

\newcommand{\ZZ}{\mathbb{Z}_2}

\def\V{\mathbb{V}}

\def\tr{ \mbox{tr}}

\begin{document}

\title{
Fermionic multi-scale entanglement renormalization ansatz}

\author{Philippe Corboz}
\affiliation{School of Physical Sciences, the University of
Queensland, QLD 4072, Australia} 
\author{Guifr\'e Vidal} 
\affiliation{School of Physical Sciences, the University of
Queensland, QLD 4072, Australia} 

\date{\today}

\begin{abstract}

In a recent contribution [arXiv:0904:4151] entanglement renormalization was generalized to fermionic lattice systems in two spatial dimensions. Entanglement renormalization is a real-space coarse-graining transformation for lattice systems that produces a variational ansatz, the multi-scale entanglement renormalization ansatz (MERA), for the ground states of local Hamiltonians. In this paper we describe in detail the fermionic version of the MERA formalism and algorithm. Starting from the bosonic MERA, which can be regarded both as a quantum circuit or in relation to a coarse-graining transformation, we indicate how the scheme needs to be modified to simulate fermions. To confirm the validity of the approach, we present benchmark results for free and interacting fermions on a square lattice with sizes between $6 \times 6$ and $162\times 162$ and with periodic boundary conditions. The present formulation of the approach applies to generic tensor network algorithms.
\end{abstract}


\maketitle

\section{Introduction}
The simulation of strongly correlated fermions in two dimensions remains one of the biggest challenges in computational physics. Quantum Monte Carlo is very powerful in solving (unfrustrated) bosonic problems, but it fails for fermionic systems because of the \textit{negative sign problem},\cite{sign} which implies an exponential scaling of the computational cost with system size and inverse temperature. Accurate simulations of fermions are crucial to gain further insight into phenomena where strong correlations play an essential role, such as high-temperature superconductivity and the fractional quantum Hall effect.   
Progress in this direction has been made in recent years with various other methods such as the cluster dynamical mean-field theory,\cite{Maier05} variational Monte Carlo,\cite{Sorella02} Gaussian Monte Carlo,\cite{Corney04} and diagrammatic Monte Carlo.\cite{Prokofev98} However, even the phase diagram of the simplest lattice model of strongly correlated electrons, the Hubbard model,\cite{Hubbard63} is still controversial. 

One dimensional fermionic problems can be accurately solved by the successful density matrix renormalization group (DMRG) method.\cite{Noack95} However, DMRG-type approaches fail for large systems in two dimensions because of an accumulation of short-range entanglement across block boundaries under successive renormalization group (RG) transformations. In recent years several ideas to extend DMRG to higher dimensions  by means of \textit{tensor networks} have been developed.\cite{Sierra98,Nishino98,Nishio04,Verstraete04,Verstraete07,MERA,Jordan08,Gu08,Jiang08}  We focus here on one particular class of tensor networks called the multi-scale entanglement renormalization ansatz (MERA), which is based on the concept of \emph{entanglement renormalization}.\cite{ER} The key idea is to apply unitary transformations (\textit{disentanglers})  locally to the system in order to remove short-range entanglement  before each coarse-graining step. This prevents the accumulation of degrees of freedom under successive RG transformations, so that arbitrarily large lattice sizes for critical and non-critical systems in one and two dimensions can be addressed.  
The MERA is a variational ansatz of the ground-state (or low energy subspace) of a system, from which arbitrary local observables and two-point correlators can be easily extracted. The accuracy of the ansatz depends on the amount of entanglement in the system, and can be controlled by a refinement parameter $\chi$.
In one-dimensional lattices, the scheme has been used to study several quantum spin systems \cite{ER,algorithm,Pfeifer09} and shown to be particularly suited to study quantum critical points \cite{ER, MERA, FreeFermions,FreeBosons,Giovannetti08, Pfeifer09}. In two dimensions, accurate results have been obtained for free fermionic and bosonic systems, \cite{FreeFermions, FreeBosons} as well as quantum spin systems,\cite{Cincio08, 2D, Evenbly09} including large lattices beyond the reach of exact diagonalization and DMRG,\cite{2D} and frustrated antiferromagnets beyond the reach of quantum Monte Carlo.\cite{Evenbly09}  In addition, an analytical MERA characterization has been provided for the ground states of a large class of models with topological order.\cite{Aguado08, Koenig08}

In a recent paper,\cite{Corboz09} entanglement renormalization and the MERA were generalized to fermionic systems. Here we present a more detailed description of the fermionic MERA and provide additional benchmarking results for free and interacting fermions in two dimensional lattices. The paper is organized as follows. In Sec. \ref{sec:bmera} we overview the MERA formalism as a means to prepare its generalization to fermionic systems. The MERA is presented both as a quantum circuit and as implementing a coarse-graining transformation. Practical calculations involve contracting diagrams or tensor networks. These correspond to different elements (such as the \emph{ascending} and \emph{descending} superoperators and \emph{environments}) needed in order to compute expectation values from the MERA or to optimize this variational ansatz.
 
In Sec. \ref{sec:fmera} we introduce the two incredients necessary in order to represent and simulate fermions. First, the tensors that constitute the MERA, namely \emph{disentanglers} and \emph{isometries}, are chosen to be parity invariant or $\mathbb{Z}_2$ symmetric. $\mathbb{Z}_2$ symmetric tensors are convenient in order to account for parity preservation. Second, we associate a \emph{fermionic swap gate} to every crossing of lines in a diagram. This gate accounts for fermionic statistics, and is the key ingredient that distinguishes the bosonic and fermionic MERA approaches. Remarkably, the cost of simulations does not depend on the particle statistics, but only on the amount of entanglement in the system.
 
Sec. \ref{sec:results} presents benchmark results. First a small lattice made of $6\times 6$ sites is analyzed with the fermionic MERA and a fermionic tree tensor network (TTN), to confirm that the ansatz can accurately represent ground states. Both free and interacting systems are analyzed. Then much larger lattices, with up to $162\times 162$ sites and periodic boundary conditions, are addressed in order to demonstrate the scalability of the present approach. Finally, Sec. \ref{sec:conclusion} contains some conclusions, and the appendices A, B and C provide some additional details.
 
The present approach to account for fermions in a tensor network is equivalent to the one presented in Ref. \onlinecite{Corboz09}. We comment on this equivalence in appendix \ref{sec:ea}. The present form of the approach, however, makes its generalization to other tensor network algorithms, such as PEPS (see also Ref. \onlinecite{fPEPS}) straightforward, as illustrated in Ref. \onlinecite{FCTM}.


\section{Bosonic MERA revisited}
\label{sec:bmera}
\subsection{Quantum circuit and renormalization group transformation}
Consider a lattice ${\cal L}_0$ of $N$ sites, where each site is described by a local Hilbert space $\mathbb{V}_0$ of finite dimension $d$. The MERA is an ansatz to describe certain states $\ket{\Psi}$ of the total Hilbert space  $\mathbb{V}_{{\cal L}_0} \cong \mathbb{V}_0^{\otimes N}$, such as the ground state of a local Hamiltonian. The ansatz is \textit{efficient} in the sense that the number of parameters required to encode a state of a translation invariant system is only of order $O(log(N) \chi^q)$, with $\chi$ a refinement parameter (see below) and $q$ a small integer number. This is in contrast to the dimension $d^N$ of the Hilbert space which grows exponentially with $N$.

\begin{figure}[!htb]
\begin{center}
\includegraphics[width=8cm]{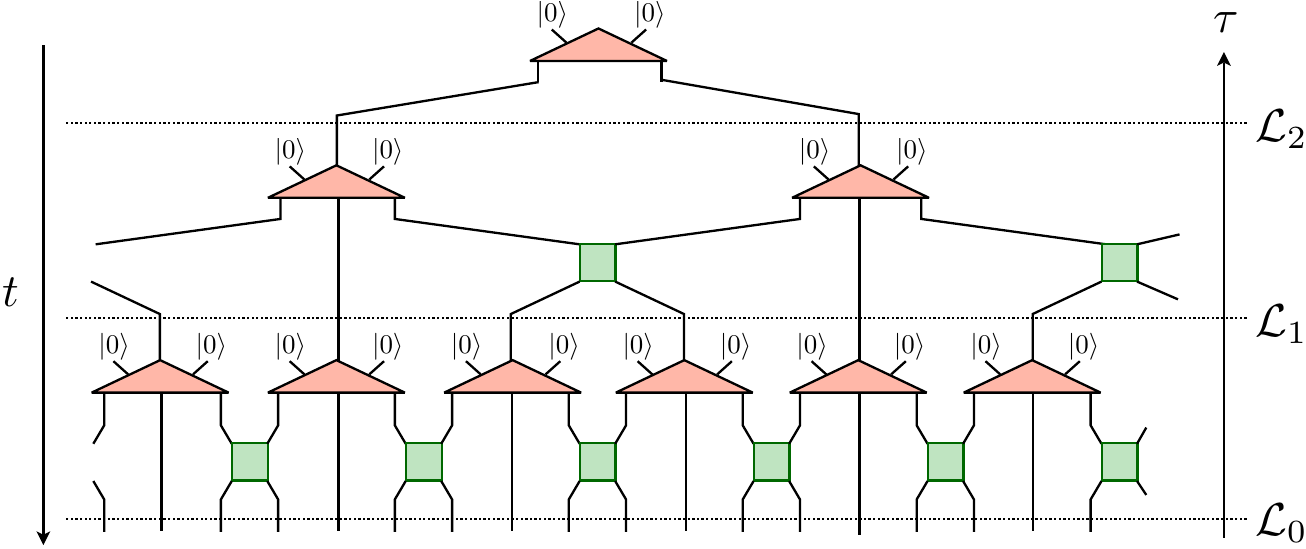}
\caption{(Color online) 
The 1D MERA represented as a quantum circuit. It consists of different types of isometric gates: disentanglers (squares) and isometries (triangles). From the perspective of a renormalization group transformation the lattice ${\cal L}_{\tau -1}$ is mapped to a coarse-grained lattice ${\cal L}_{\tau}$ by applying a layer of disentanglers and isometries (cf. Fig. \ref{fig:mera1D}). } 
\label{fig:qcircuit}
\end{center}
\end{figure}

The MERA can be regarded as a quantum circuit whose output wires correspond to the sites of the lattice ${\cal L}_0$ as depicted in Fig. \ref{fig:qcircuit}. We first focus on the ternary 1D MERA scheme introduced in Ref. \onlinecite{algorithm}, and then on its generalization to the 2D case. Several unitary gates transform the untentangled state $\ket{0}^{\otimes N}$ into a state $\ket{\Psi} \in \mathbb{V}_{{\cal L}_0} $. We distinguish between two types of gates, \textit{isometries} $w$ and \textit{disentanglers} $u$, each only involving a small number of input and output wires. A disentangler $u$ is a map
\begin{equation}
\label{eq:u}
u : \mathbb{V}^{\otimes  2} \rightarrow \mathbb{V}^{\otimes 2}, \quad u^{\dagger} u=u u^{\dagger} = I_{\V^{\otimes 2}},
\end{equation}
with $I_{\V^{\otimes 2}}$ the identity operator in  $\mathbb{V}^{\otimes 2}$, and 
an isometry $w$ is a map
\begin{equation}
\label{eq:w}
w : \V  \rightarrow \V^{\otimes 3}, \quad w^{\dagger} w = I_{ \V},
\end{equation}
with $I_{ \V}$ the identity operator in $ \V$. 

\begin{figure}[!htb]
\begin{center}
\includegraphics[width=8cm]{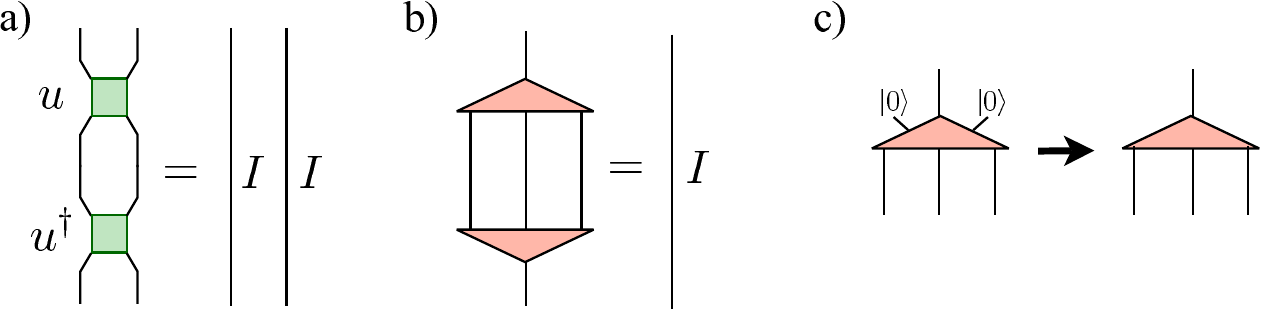}
\caption{(Color online) 
Disentanglers (a) and isometries (b) in the MERA are isometric gates. c) We usually draw the isometry without the incoming wires that are fixed to $\ket{0}$.  } 
\label{fig:gates}
\end{center}
\end{figure}

\begin{figure}[!htb]
\begin{center}
\includegraphics[width=8cm]{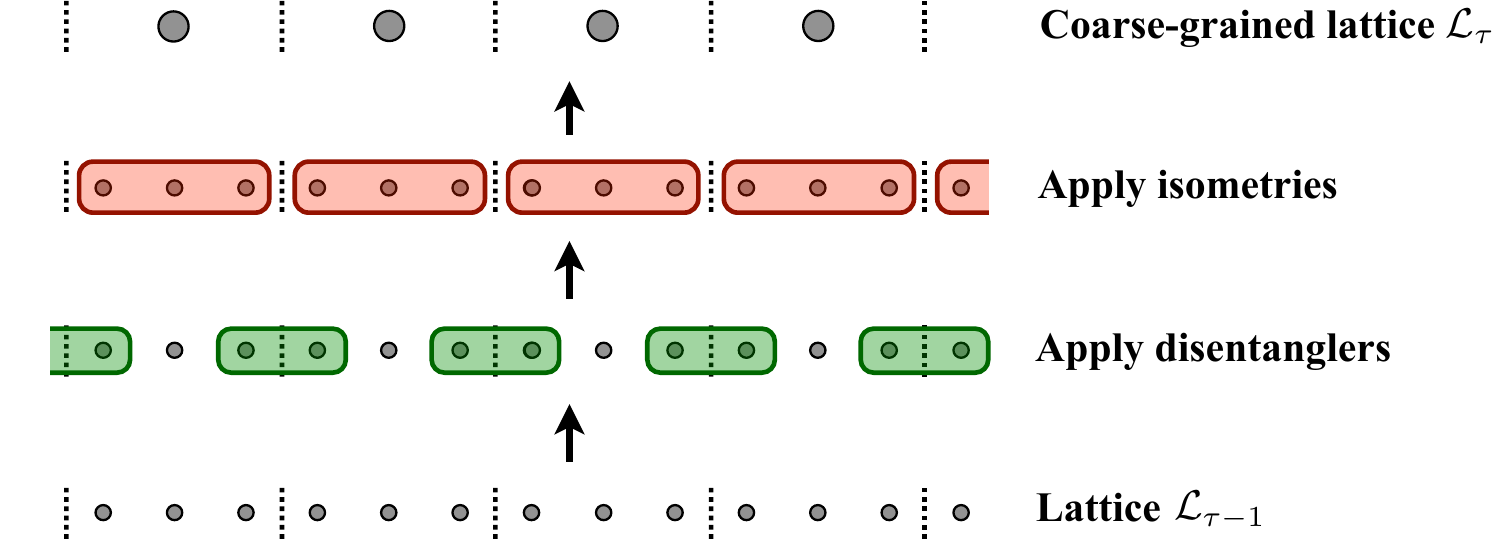}
\caption{(Color online) 
The real-space renormalization group transformation of the 1D MERA (ternary scheme). } 
\label{fig:mera1D}
\end{center}
\end{figure}

From the perspective of a renormalization group transformation (see Ref. \onlinecite{ER}), an isometry coarse-grains three sites into one effective site. 
A disentangler $u$ acts across the boundary of two blocks of sites to reduce the amount of short-range entanglement between the blocks.\cite{ER} The application of a layer of disentanglers followed by a layer of isometries describes a mapping of the lattice ${\cal L}_{\tau-1}$ into a coarse-grained lattice ${\cal L}_{\tau}$, as shown in Fig. \ref{fig:mera1D}. The local dimension of each coarse-grained site is denoted by $\chi$.

Another key feature of the MERA is its causal structure. The past \textit{causal cone} of an outgoing wire $s$ at time $t$ is defined as the set of gates and wires that can affect the state in $(s,t)$. A MERA is a quantum circuit for which the past causal cone of any location $(s,t)$ in the circuit has a \textit{bounded width}, i.e. involves only a constant number (independent of $N$) of wires at any previous time $t'<t$.

\begin{figure}[!htb]
\begin{center}
\includegraphics[width=8cm]{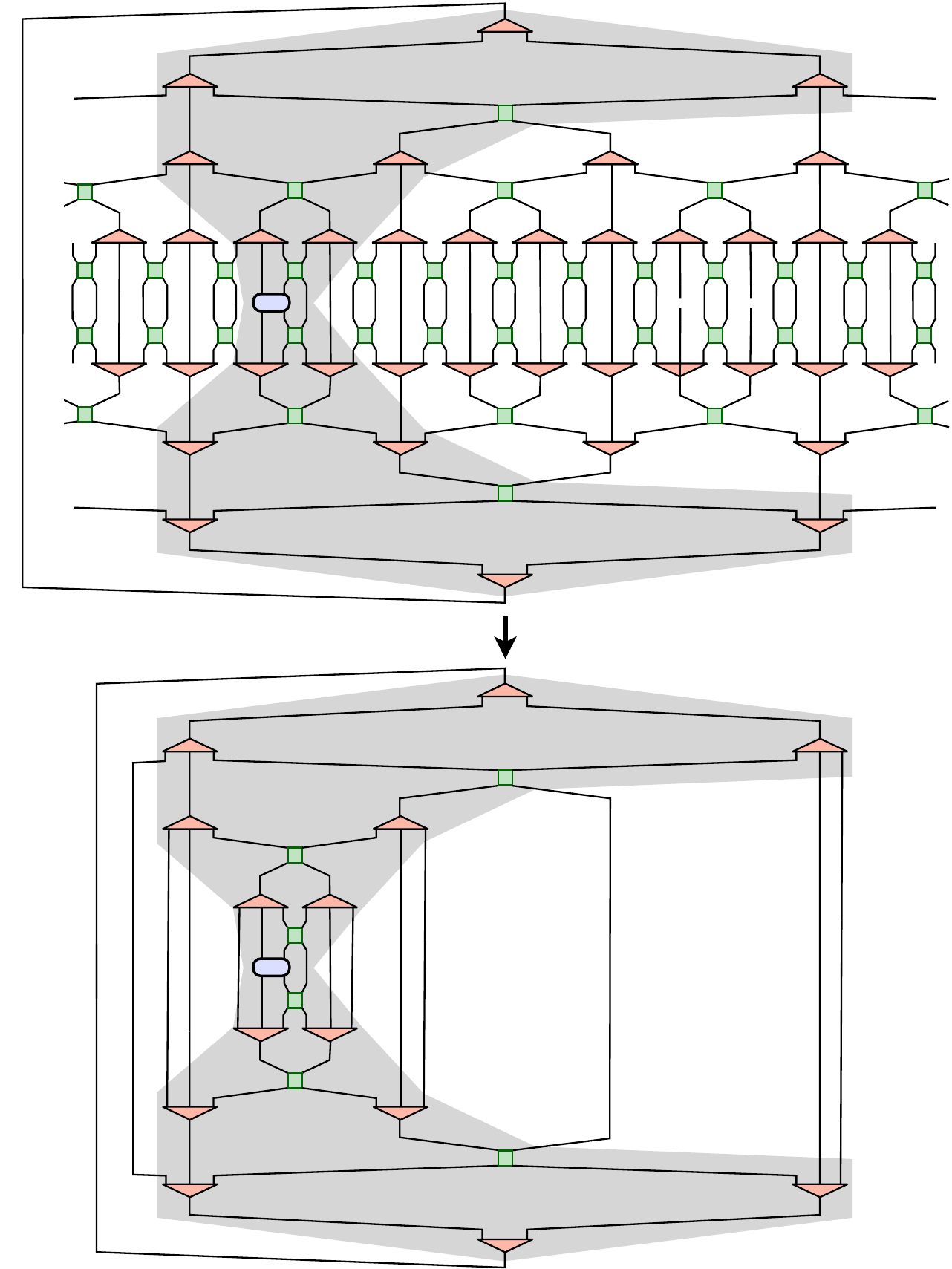}
\caption{(Color online) 
\textit{Top}: The causal cone of an operator (oval) given by the shaded area involves only a small number of gates. \textit{Bottom}: All gates outside the causal cone annihilate and we are left with a much simpler circuit.} 
\label{fig:causalcone}
\end{center}
\end{figure}

These key properties of the MERA enable an efficient calculation of expectation values of local observables, $\expect{\Psi | \hO | \Psi}$, since only gates included in the causal cone of the operator $\hO$ (i.e. the causal cones of the wires connected to $\hO$) have to be taken into account. All other gates can be replaced by identity operators thanks to Eqs. \eqref{eq:u} and \eqref{eq:w}, as illustrated in Fig. \ref{fig:causalcone}.

\subsection{Superoperators and environments}
\begin{figure}[!htb]
\begin{center}
\includegraphics[width=7.9cm]{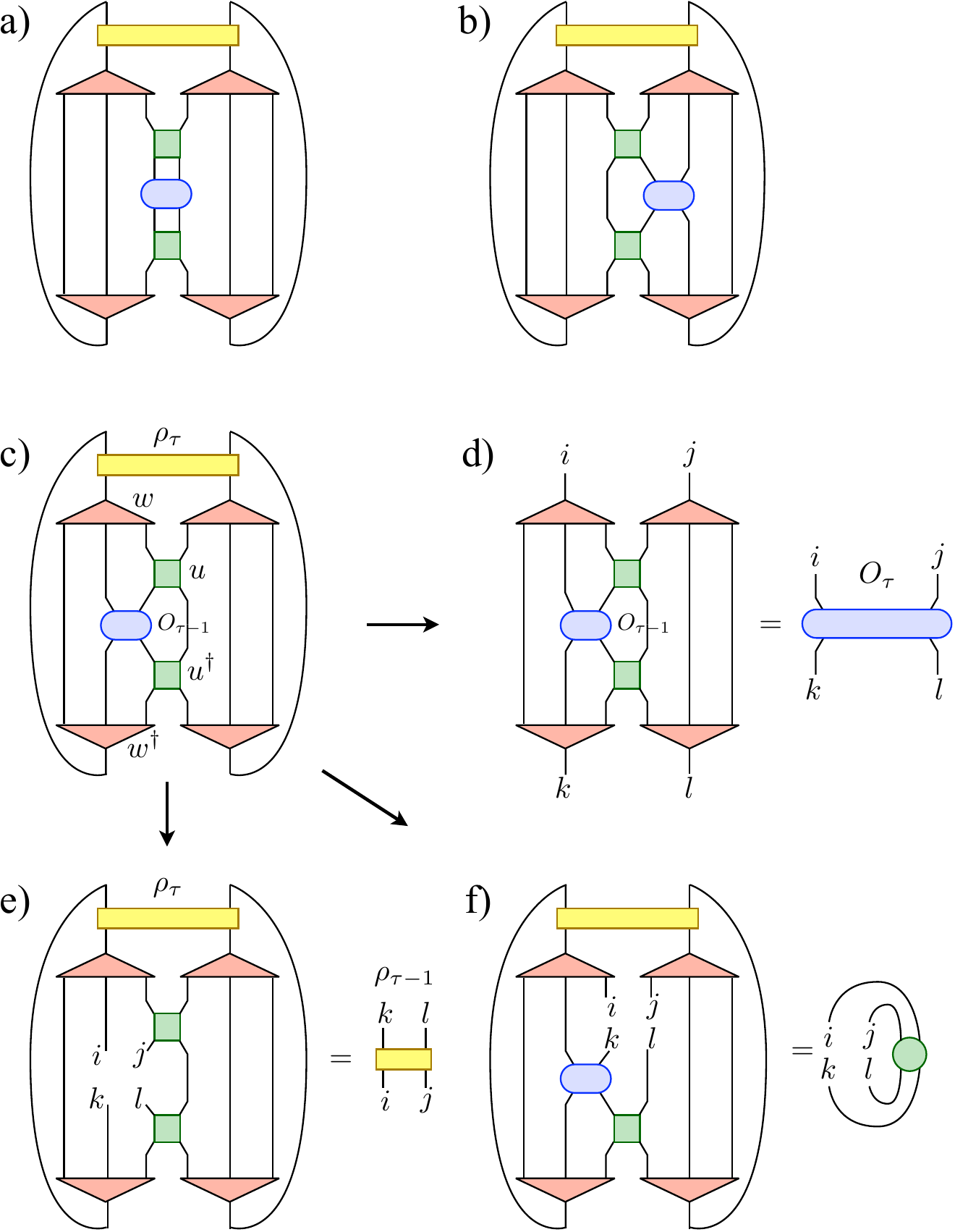}
\caption{(Color online) 
a)-c) The three generating diagrams of the 1D MERA. d) An ascending diagram resulting from c) by erasing the two-site density matrix $\rho_\tau$. e) A descending diagram obtained from c) by taking away the operator $O_{\tau-1}$. f) An environment for the disentangler $u$ created by erasing the (upper) disentangler from c). An environment for $w$ can be obtained in a similar way.}
\label{fig:diagrams1D}
\end{center}
\end{figure}

It has been shown in Ref. \onlinecite{algorithm} that the systematic evaluation and manipulation of a MERA boils down to the calculation of several small diagrams which fall into three classes: \textit{ascending superoperators}, \textit{descending  superoperators}, and \textit{environments}. All these diagrams can be constructed from the three \textit{generating diagrams} shown in Fig. \ref{fig:diagrams1D}a)-c), as we explain in the following.  

\textit{Ascending superoperators - }
An ascending superoperator $\cal A$ transforms a two-site operator $O_{\tau-1}$ defined on lattice ${\cal L}_{\tau-1}$ into a two-site operator $O_{\tau}$ on the coarse-grained lattice $\cal L_{\tau}$, as shown in Fig. \ref{fig:diagrams1D}d). There are three structurally different ascending superoperators, which result from the three generating diagrams in Fig. \ref{fig:diagrams1D}a)-c) by taking away the density matrix $\rho_\tau$ from each diagram.  Repeated application of the corresponding ascending superoperator to $O_{0}$ creates a sequence of increasingly coarse-grained operators $\{O_0, O_1, ..., O_T\}$.

\textit{Descending superoperators - }
A descending superoperator $\cal D$ maps a two-site density matrix $\rho_\tau$ on the lattice ${\cal L}_{\tau}$ into a two-site density matrix on the finer (less coarse-grained) lattice ${\cal L}_{\tau-1}$, as illustrated in Fig. \ref{fig:diagrams1D}e). The three different descending superoperators can be obtained from the generating diagrams by erasing the operator $O_{\tau-1}$ from each diagram. Repeated application of the corresponding descending superoperator to $\rho_{T-1}$ creates a sequence of increasingly finer two-site density matrices $\{\rho_{T-1},\rho_{T-2}, ..., \rho_0\}$. Note that $\rho_{T-1}$ is obtained by joining the top-isometry with its conjugate.  

\textit{Environments - }
There are several ways to optimize the MERA. In this work we used the algorithm from Ref. \onlinecite{algorithm}, which is based on iterative optimization of individual gates. The optimization of, e.g., a disentangler involves calculating its three different environments, which are obtained by erasing the (upper) disentangler from the generating diagrams, as shown for example in Fig. \ref{fig:diagrams1D}f). Note that an isometry has six different environments, three resulting from erasing the left isometry from the generating diagrams, and the other three from erasing the one on the right (see Ref. \onlinecite{algorithm} for more details).

\subsection{Diagrams}
\label{sec:objects}
\begin{figure}[!htb]
\begin{center}
\includegraphics[width=8cm]{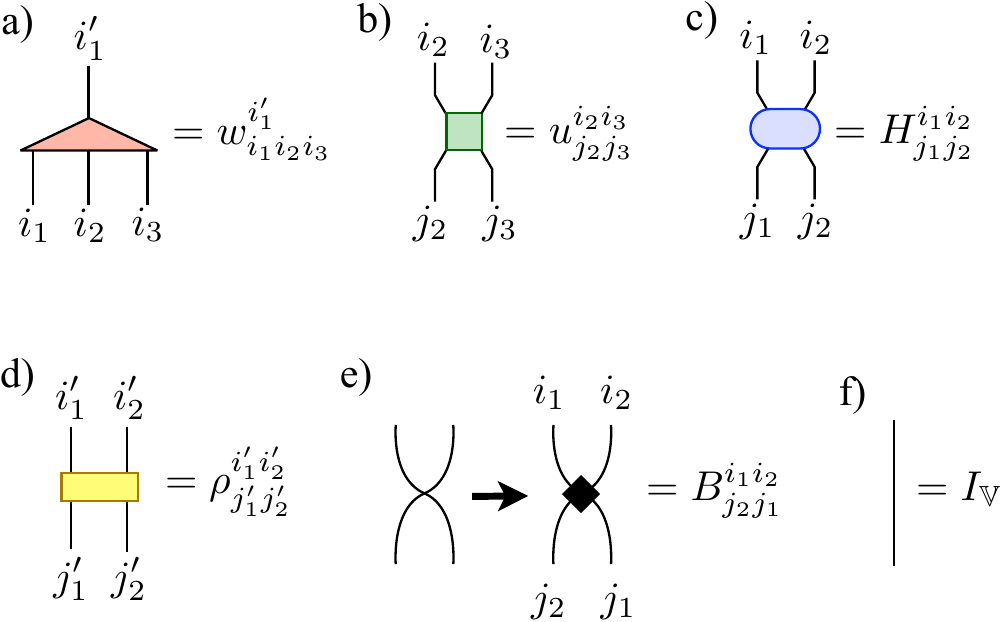}
\caption{(Color online) 
Elements in a diagram of the 1D MERA. Each shape represents a tensor: a) isometry, b) disentangler, c) Hamiltonian, d) density matrix. e) A crossing of lines corresponds to a two-body gate (which is simply the identity in the bosonic MERA). f) A single line corresponds to the identity $I_\V$ of the vector space $\V$.
} 
\label{fig:objects}
\end{center}
\end{figure}

Once we determined the diagrammatic representation of the ascending/descending superoperators and environments we can evaluate the diagram by contracting the corresponding tensor network. We start by identifying all elements appearing in a diagram, which are summarized in Fig. \ref{fig:objects}.

\subsubsection{Elements in a diagram}
\textit{Shapes - }
Each shape represents a tensor (a multidimensional array) with a rank equal to the number of legs. The entries of a tensor are given by the expansion coefficients of the corresponding gate (or Hamiltonian/density matrix) in the local bases of its legs.
For example, a general two-body Hamiltonian term can be expanded as 
\begin{equation}
{\hat H} = \sum_{i_1 i_2 \atop j_1 j_2} H^{i_1 i_2}_{j_1 j_2} \ket{j_1 j_2} \bra{i_1 i_2}
\end{equation}
where each sum goes over all basis states of the local Hilbert space of each leg. The four-leg tensor associated to $\hH$ is $H^{i_1 i_2}_{j_1 j_2}$. 

\textit{Line crossings -}
\label{sec:crossings}
A diagram may involve line crossings, e.g. the 1D MERA in the case of periodic boundary conditions (or for a Hamiltonian with next-nearest neighbor interaction), or the 2D MERA as we will see in Sec. \ref{sec:mera2D}. Each crossing corresponds to an exchange of the degrees of freedom carried by the individual lines. The implications of this exchange depend on the statistics of the basic degrees of freedom. In general, we replace each crossing by a \textit{swap gate} $B$ which accounts for the exchange process (see appendix \ref{app:swap}). In the bosonic MERA this gate is simply the identity, i.e. $B^{i_1 i_2}_{j_2 j_1}=\delta_{i_1 j_1} \delta_{i_2 j_2}$, because the bosonic wavefunction is symmetric under exchange of two particles. As a consequence the crossings can simply be ignored. However, this will no longer hold in the fermionic MERA as we will see Sec. \ref{sec:fmera}!

\textit{Lines - }
A single line in a diagram corresponds to the identity $I_\V$ of the vector space $\V$. A line connecting two tensors describes how they are multiplied together, when the diagram is contracted, as we explain next.

\subsubsection{Contraction of a diagram}
\begin{figure}[!htb]
\begin{center}
\includegraphics[width=8cm]{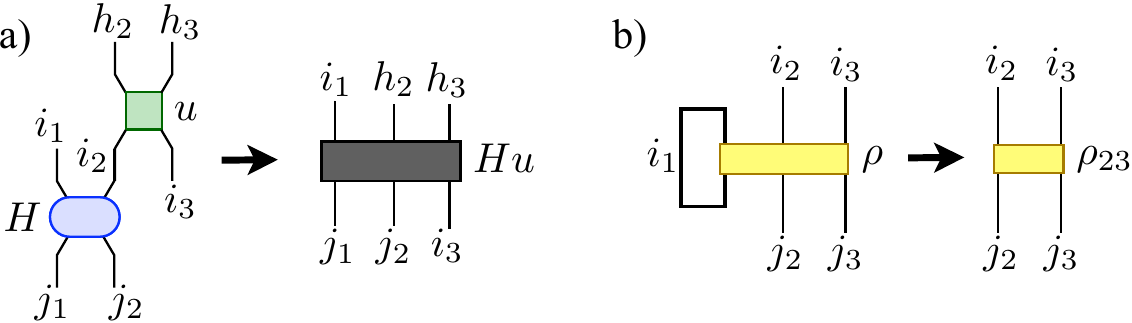}
\caption{(Color online) 
a) Multiplication of two tensors $H$ and $u$ on the legs connected by the line labelled by $i_2$. b) A line connecting two legs of the same tensor corresponds to a trace (see text). } 
\label{fig:mult}
\end{center}
\end{figure}
\begin{figure}[!htb]
\begin{center}
\includegraphics[width=8cm]{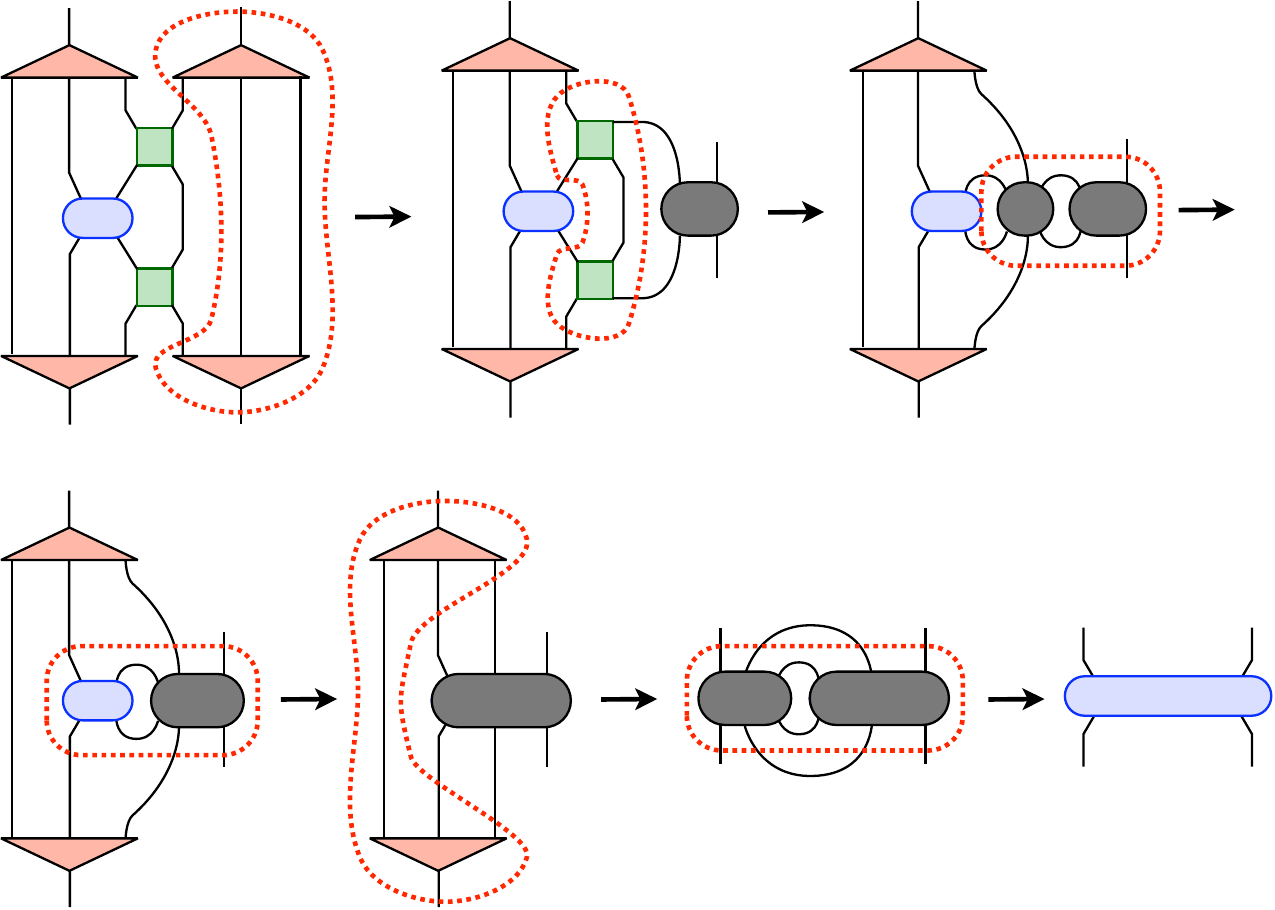}
\caption{(Color online) 
Contraction of the tensor network in Fig. \ref{fig:diagrams1D}d) by a sequence of pairwise multiplication of tensors.} 
\label{fig:contraction}
\end{center}
\end{figure}

A tensor network is contracted by a sequence of pairwise multiplication of tensors. Two tensors are multiplied together according to the lines connecting the legs of the tensors. 
For example, the multiplication in Fig. \ref{fig:mult}a) of  $H^{i_1 i_2}_{j_1 j_2}$ with $u^{h_2 h_3}_{i_2 i_3}$ on the leg labelled by $i_2$ leads to a new tensor $Hu$ given by
\begin{equation}
[Hu]^{i_1 h_2 h_3}_{j_1 j_2 i_3} = \sum_{i_2} H^{i_1 i_2}_{j_1 j_2} u^{h_2 h_3}_{i_2 i_3}.
\end{equation}
A special case is the trace, which is represented as a line connecting two legs of the same tensor, as for example
\begin{equation}
[\rho_{23}]^{i_2 i_3}_{j_2 j_3} = \sum_{i_1 }  \rho^{i_1 i_2 i_3}_{i_1 j_2 j_3}, \quad 
\end{equation}
illustrated in Fig. \ref{fig:mult}b).   
An example of a full contraction of a tensor network is shown in Fig. \ref{fig:contraction}. 

Note that the computational cost depends on the \textit{order} in which the pairwise multiplications are implemented. In a practical implementation it is therefore crucial to determine the order which minimizes the computational cost (and/or memory requirements). The computational cost to multiply two tensors $A$ and $B$ connected by $l_c$ legs, is given by $\chi^{l_{A}+l_{B}-l_c}$, where $l_{A}$ ($l_B$) is the number of legs of tensor $A$ ($B$), and we assumed that each leg has the same dimension $\chi$. The scaling of a MERA algorithm is dominated by the largest cost in the contraction of a diagram. The cost in memory scales with $\chi^{l_{max}}$, with $l_{max}$ the tensor with the biggest number of legs occurring during the contraction. For the 1D ternary MERA the computational cost scales as $O(\chi^8)$, and the cost in memory as $O(\chi^6)$.\cite{comment:d}

\subsection{2D MERA}
\label{sec:mera2D}
\begin{figure}[!htb]
\begin{center}
\includegraphics[width=8cm]{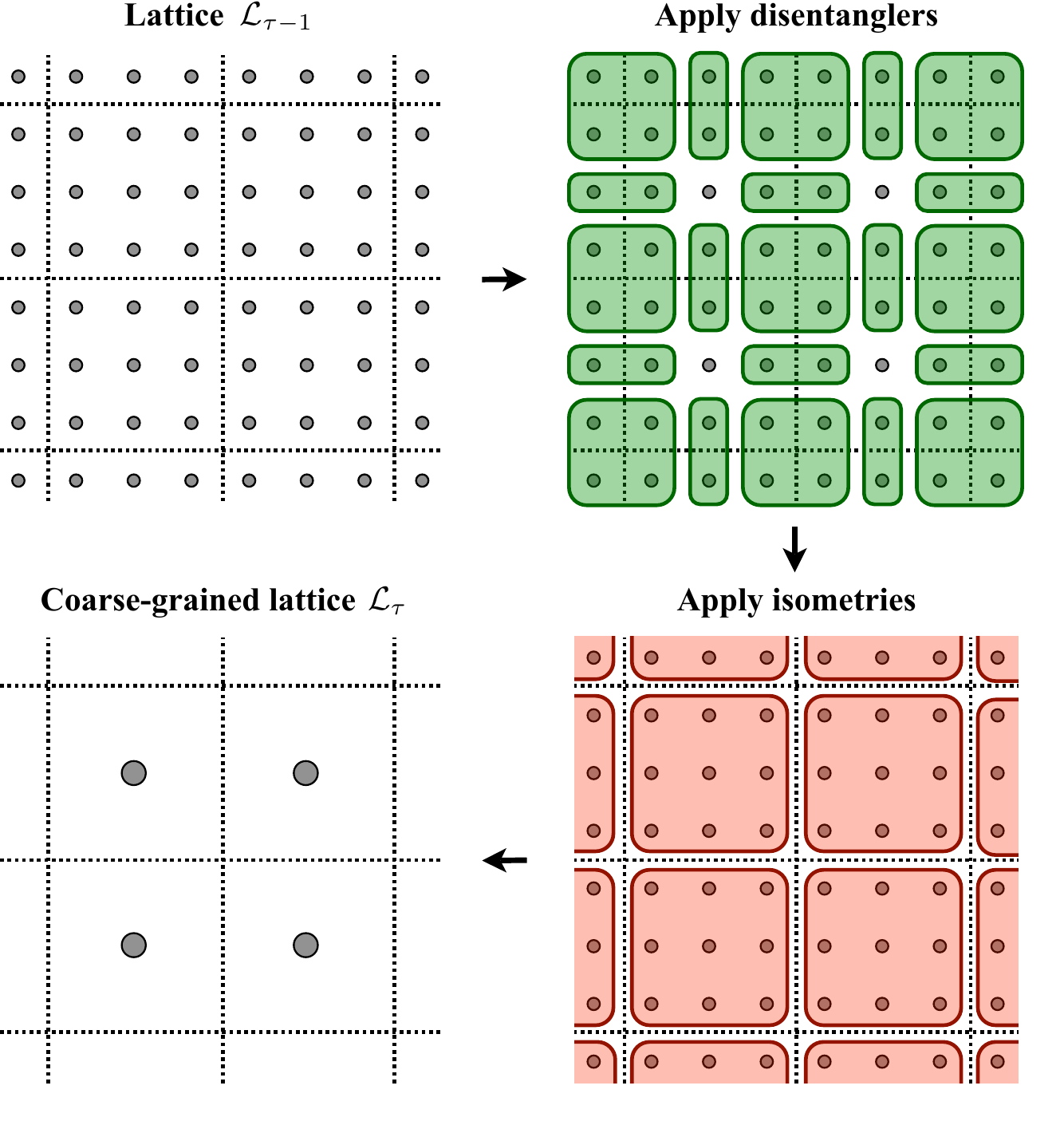}
\caption{(Color online) 
The real-space renormalization group transformation of the 2D MERA.} 
\label{fig:mera2D}
\end{center}
\end{figure}

There are several ways to realize a MERA in 2D.\cite{MERA, FreeFermions,  Cincio08, 2D, algorithm} Here we focus on a "9-to-1 scheme" where a site of $\cal L_\tau$ corresponds to a block of $3\times 3=9$ sites of  ${\cal L}_{\tau-1}$, and with disentanglers that do not overlap, as depicted in Fig. \ref{fig:mera2D}. The computational cost of this scheme scales as $O(\chi^{16})$, and the cost in memory as $O(\chi^{12})$.

Conceptually, one proceeds in the same way as in the 1D MERA, i.e. one determines ascending/descending superoperators and environments.  The ascending superoperator maps a 4-body plaquette operator $O_{\tau-1}$ on the lattice ${\cal L}_{\tau-1}$ into a plaquette operator $O_{\tau}$ on the lattice $\cal L_{\tau}$. But a 2-body operator, depending on its location in the lattice, may be mapped into a 4-body operator on a higher level. We therefore focus here on plaquette operators. (Note that in case of a two-body Hamiltonian we can either treat the lowest layer differently than the higher layers, or express the hamiltonian as a sum of plaquette operators from the start).
%

\begin{figure}[!htb]
\begin{center}
\includegraphics[width=8cm]{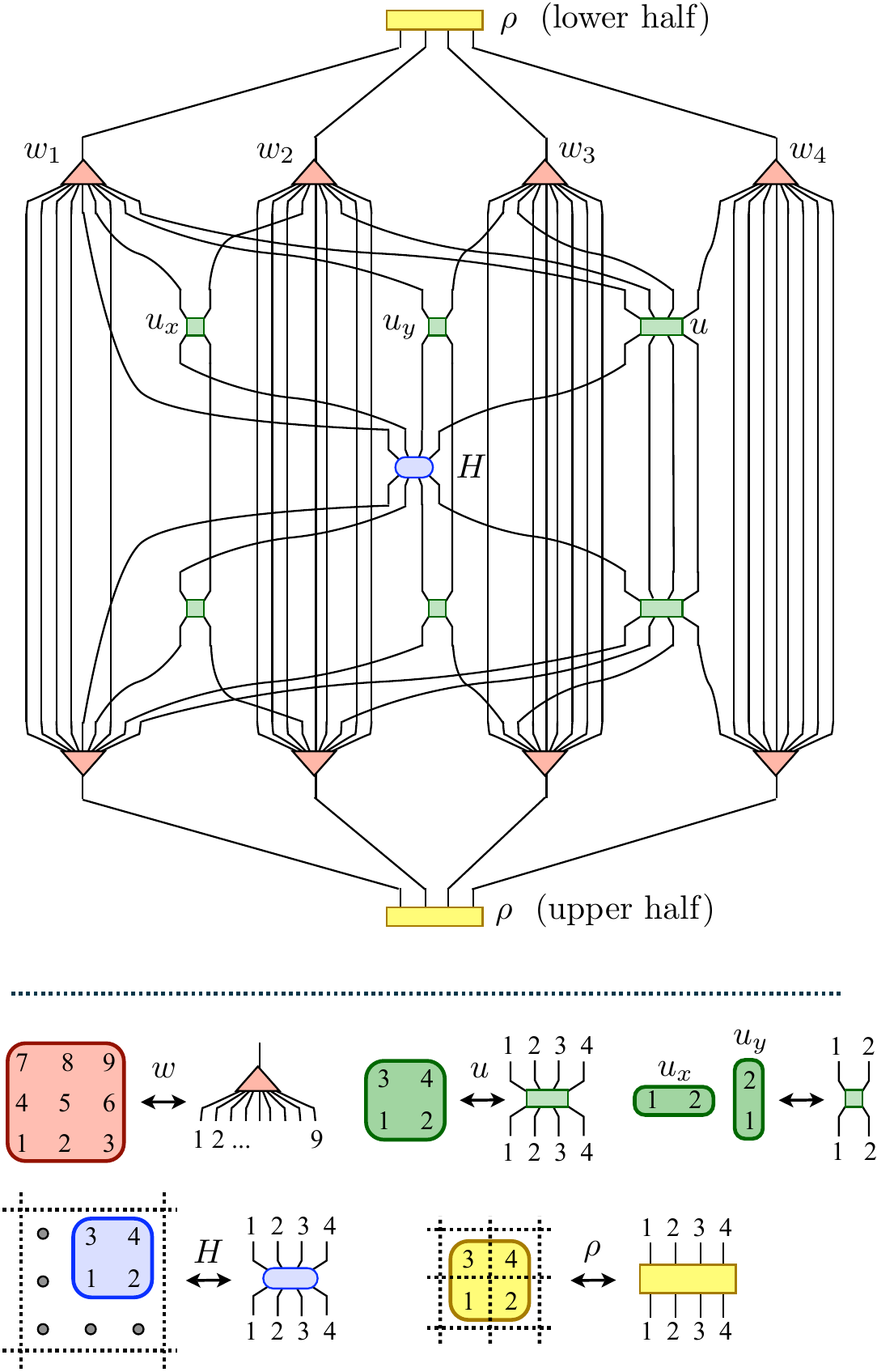}
\caption{(Color online) 
Example of a generating diagram of the 2D MERA projected onto 1+1 dimensions. On the bottom we define the correspondence between the gates in this figure with the ones from Fig. \ref{fig:mera2D}, as indicated by the numbers.  The picture on the bottom left shows the location of the 4-body plaquette Hamiltonian (oval) in the lattice. }
\label{fig:diagram2D}
\end{center}
\end{figure}

Figure \ref{fig:diagram2D} shows one particular generating diagram, from which we can obtain ascending/descending superoperators or an environment by eliminating the corresponding tensor, as explained for the 1D case. 
There are 9 different generating diagrams, corresponding to the 9 different positions of the Hamiltonian with respect to the basic $3 \times 3$ block. 

Note that the basic diagrams of a 2D MERA are $2+1$ dimensional objects. In Figure \ref{fig:diagram2D} we chose one particular way to map the diagram onto a plane, i.e. 1+1 dimensions like the 1D MERA. This mapping is not unique, but for any choice, some of the lines in the diagram cross each other. As already mentioned, these crossings can be ignored in the bosonic case. However, they will play an important role for the fermionic MERA, as we will explain in the next section.

\section{Fermionic MERA}
\label{sec:fmera}
The essential difference between a fermionic and a bosonic system lies in the symmetry of the wavefunction under the exchange of two particles. Exchanging two bosons leaves the wavefunction invariant, whereas when exchanging two fermions the wavefunction is multiplied by $-1$. 
More generally, exchanging an odd number of fermions living on a (coarse-grained) site $i'$ with an odd number of fermions on $j'$ leads to a negative sign.

All basic concepts introduced for the bosonic MERA still hold for the fermionic MERA, i.e. the gates are isometric and the causal cone is the same as in the bosonic MERA (see appendix \ref{app:cc}). All we need to do is to use parity preserving tensors, and introduce a fermionic swap gate, which implements the fermionic exchange properties, as we explain in the following.

\subsection{$\mathbb{Z}_2$ symmetry}
A property of any fermionic Hamiltonian $\hH$ (and more generally any fermionic observable) is that it \textit{preserves parity}, i.e. $[ \hP, \hH]=0$, with $\hP=(-1)^{\hat N}$ the total parity operator, where $\hat N$ measures the total number of particles in the system. This $\mathbb{Z}_2$ symmetry stems from the fact that  fermions can only be created or annihilated in pairs. We incorporate this symmetry into the MERA by enforcing all tensors to be parity preserving. A tensor $T_{i_1 i_2 \dots i_M}$ preserves parity if 
\begin{equation}
T_{i_1 i_2 \dots i_M} = 0, \quad \text{if} \,\, P(i_1) P(i_2) \dots P(i_M) \neq 1,
\end{equation}
where $P(i_k) \in \{-1,1\}$ denotes the parity of the state labelled by $i_k$.  
The local Hilbert space of a (coarse-grained) site is decomposed into a space with even parity $(+)$ and one with odd parity $(-)$, i.e. $\V = \V^{(+)} \oplus \V^{(-)}$. Each basis state in $\mathbb{V}$ is labeled now by a composite index ${j}= (p,\alpha^p)$, where $p \in \{+,-\}$ specifies the parity sector and $\alpha^p$ enumerates the states in the subspace $\mathbb{V}^{(p)}$. This decomposition allows us to identify the parity of a state very easily, and it also leads to a block structure of the tensors (similarly to a block diagonal matrix).

\begin{figure}[!htb]
\begin{center}
\includegraphics[width=8cm]{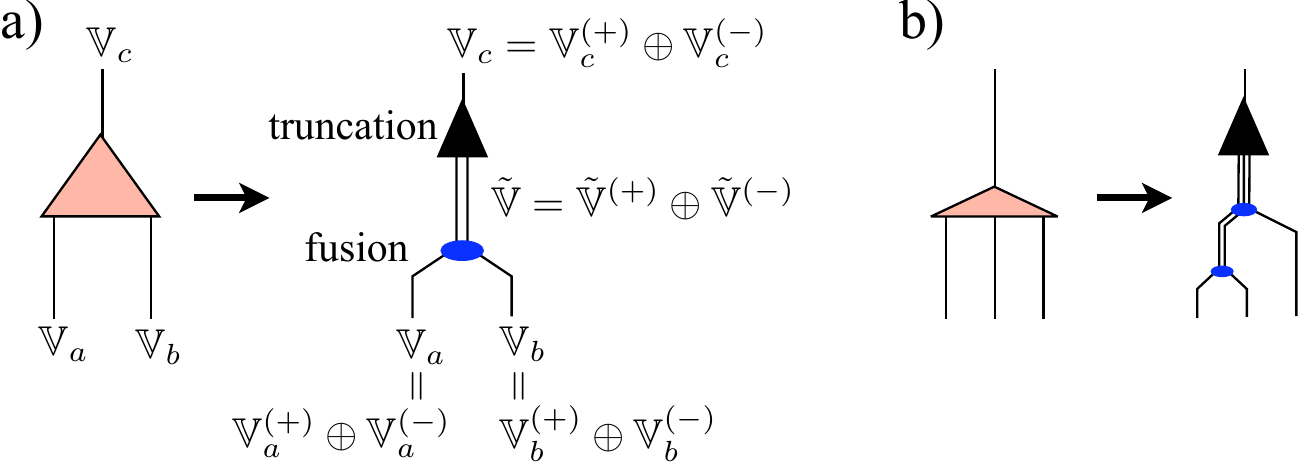}
\caption{(Color online) 
a) An isometry that coarse-grains two sites $a$ and $b$ into one site $c$ can be decomposed into a fusion of the two sites followed by a truncation of the combined Hilbert space $\tilde \V$. b) An isometry that coarse-grains three sites into one can be decomposed into two subsequent fusions, followed by a truncation. }
\label{fig:Z2isometry}
\end{center}
\end{figure}

\textit{Fusion rules -}
An isometry that coarse-grains two sites $a$ and $b$ into one site $c$ can be split into a \textit{fusion} of the two sites (blocking) followed by a \textit{truncation} of the combined Hilbert space $\tilde V = \V_a \otimes \V_b$, as illustrated in Fig. \ref{fig:Z2isometry}. 
The fusion rules  describe how the individual sectors of $\tilde \V= {\tilde \V}^{(+)} \oplus {\tilde \V}^{(-)}$ result from combining the sectors of $\V_a$ and $\V_b$:
\begin{eqnarray}
\tilde \V^{(+)} &=&  (\V^{(+)}_{a} \otimes \V^{(+)}_{b}) \oplus  (\V^{(-)}_{a} \otimes \V^{(-)}_{b}), \\
\tilde \V^{(-)} &=&  (\V^{(-)}_{a} \otimes \V^{(+)}_{b}) \oplus  (\V^{(+)}_{a} \otimes \V^{(-)}_{b}).
\end{eqnarray} 
Note that the truncation is performed separately in each sector, $\tilde \V^{(+)} \rightarrow \V^{(+)}_{c}$ and $\tilde \V^{(-)} \rightarrow \V^{(-)}_{c}$. 
Finally, a fusion of $N$ sites can be decomposed into $N-1$ two-site fusions, as illustrated in Fig. \ref{fig:Z2isometry}b). 

We emphasize that also many bosonic systems exhibit a $\mathbb{Z}_2$ symmetry, which can be incorporated into the bosonic MERA in the same way.\cite{Singh09} In general, exploiting symmetries increases the efficiency of a simulation. However, for fermions, the parity symmetry also plays an important role for the implementation of the fermionic swap gate which we introduce in the next section.

\subsection{Fermionic swap gate}
\label{sec:fswap}
\begin{figure}[!htb]
\begin{center}
\includegraphics[width=8cm]{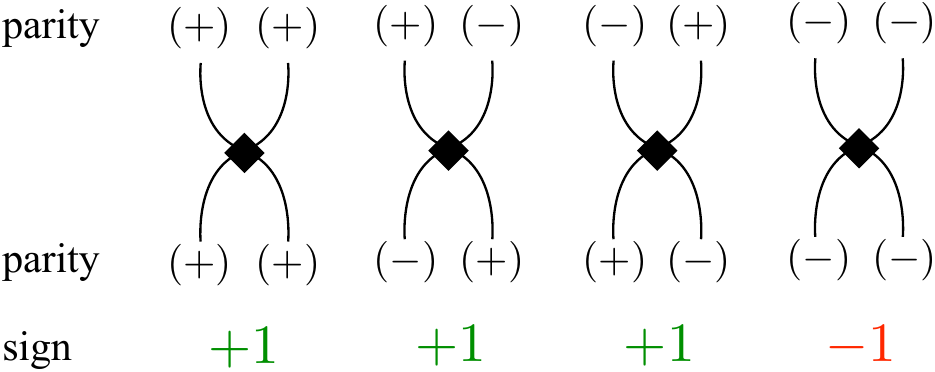}
\caption{(Color online) 
The fermionic swap gate implements an exchange of fermions. Exchanging an odd number of fermions on one site with an odd number of fermions on another site leads to a negative sign factor.  } 
\label{fig:fbraiding}
\end{center}
\end{figure}
As explained in Sec. \ref{sec:objects},
two crossing lines $i$ and $j$ in a tensor network are nothing but a graphical representation of an exchange process (or a swapping). 
As a consequence of the antisymmetry of the fermionic wavefunction, a prefactor of $-1$ appears if both lines carry a state with odd parity (odd number of particles), as illustrated in Fig. \ref{fig:fbraiding}.
We replace each crossing by a gate $B$ that accounts for this exchange process (see appendix \ref{app:swap}):
\begin{equation}
\label{eq:swapgate}
B^{i_1 i_2}_{j_2 j_1} = \delta_{i_1, j_1} \delta_{i_2, j_2} S(P(i_1),P(i_2)),
\end{equation}%
with
\begin{equation}
\label{eq:swapgate}
S(P(i_1),P(i_2)) = 1- 2 \delta_{P(i_1), -1} \delta_{P(i_2), -1}
\end{equation}
only depending on the parities of the states $i_1$ and $i_2$. The function $S$ evaluates to $-1$ if both parities are odd, and $+1$ otherwise.

Having parity preserving tensors allows us to take a line and "jump" over another tensor, as illustrated in Fig. \ref{fig:jump}. We demonstrate the validity of this transformation in appendix \ref{app:jump}. Before contracting the tensor network, we rearrange the lines and tensors in such a way that each of the resulting fermionic swap gates can be absorbed into a single tensor, as shown in Fig. \ref{fig:jump}.\cite{rem:absorbing} The resulting tensor network can then be contracted in the same way as in the bosonic case. Note that the computational cost of absorbing a fermionic swap gate into a tensor with $l$ legs is only of order $\chi^l$. Therefore, this cost is subleading and the overall cost of the algorithm is essentially the same as in the bosonic case.  

In other words, we map a non-planar tensor network to a planar one (i.e. a network without line crossings) by replacing the crossings by fermionic swap gates. We can modify the resulting planar network by "jump" moves in such a way that the resulting fermionic swap gates do not increase the leading cost of a contraction, compared to the bosonic case.

The fermionic MERA presented in this paper may look different than the one introduced in Ref. \onlinecite{Corboz09}, which is based on the Jordan-Wigner transformation to map the fermionic system into a bosonic one. However, it is important to point out that the two approaches describe the same MERA (see appendix \ref{app:swap}). 

\begin{figure}[!htb]
\begin{center}
\includegraphics[width=8cm]{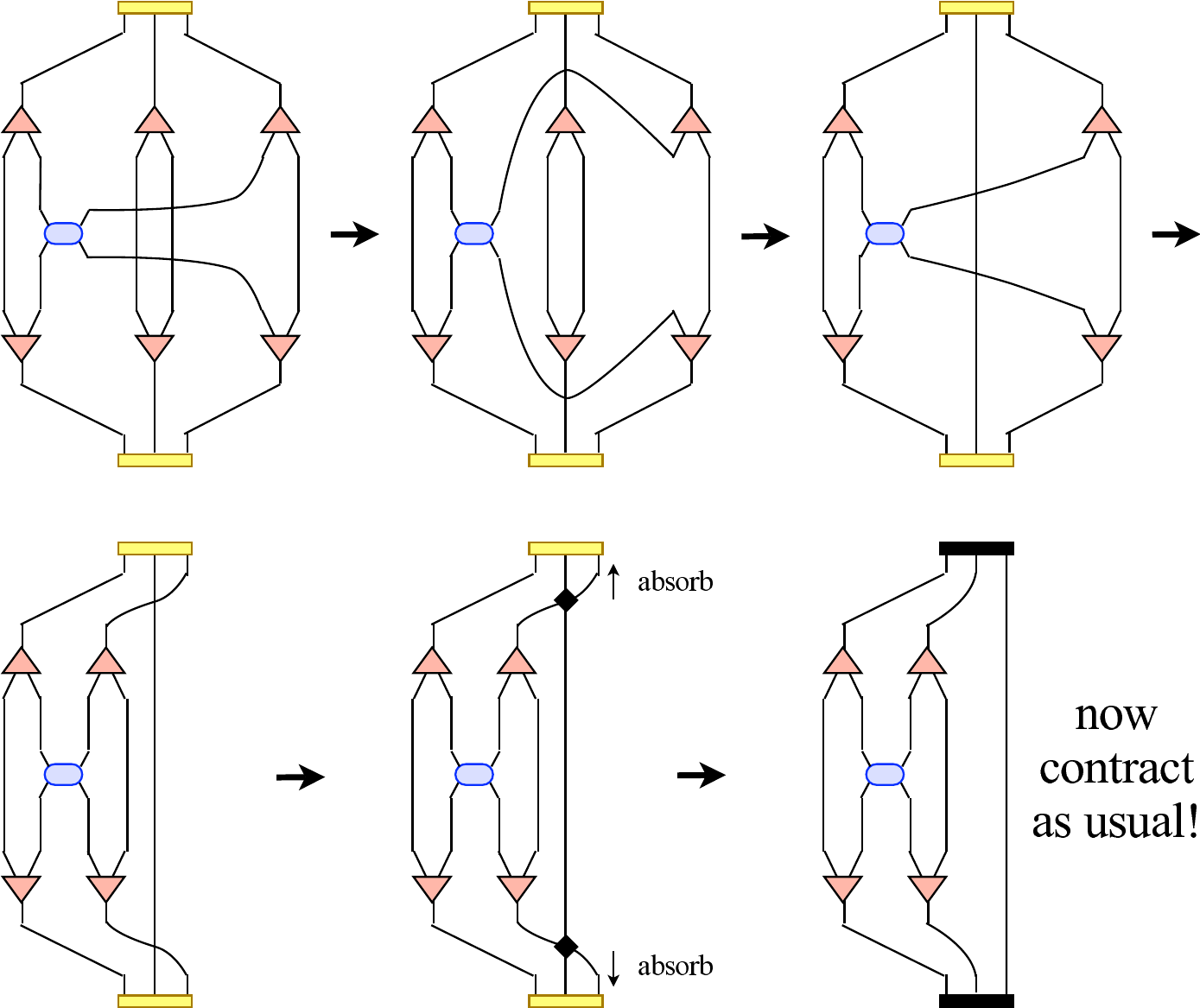}
\caption{(Color online) 
An example diagram involving line crossings. Thanks to the parity symmetry of each tensor we are allowed to "jump" with a line over a tensor in order to simplify the tensor network. Lines are moved around in such a way that each fermionic swap gate can be absorbed into a single tensor. The resulting tensor network is contracted as in the bosonic case.} 
\label{fig:jump}
\end{center}
\end{figure}

\section{Results}
\label{sec:results}
In this section we present benchmark results for the fermionic 2D MERA, and also for the 2D tree tensor network (TTN),\cite{Tagliacozzo09} which corresponds to the 2D MERA without disentanglers. Thanks to its simpler structure, a larger value of $\chi$ is affordable. However, in contrast to the MERA the TTN is not scalable, i.e. in general $\chi$ has to increase exponentially with system size to account for the accumulation of short-range entanglement across the boundary of a block. 

\begin{figure}[htb]
\begin{center}
\includegraphics[width=8cm]{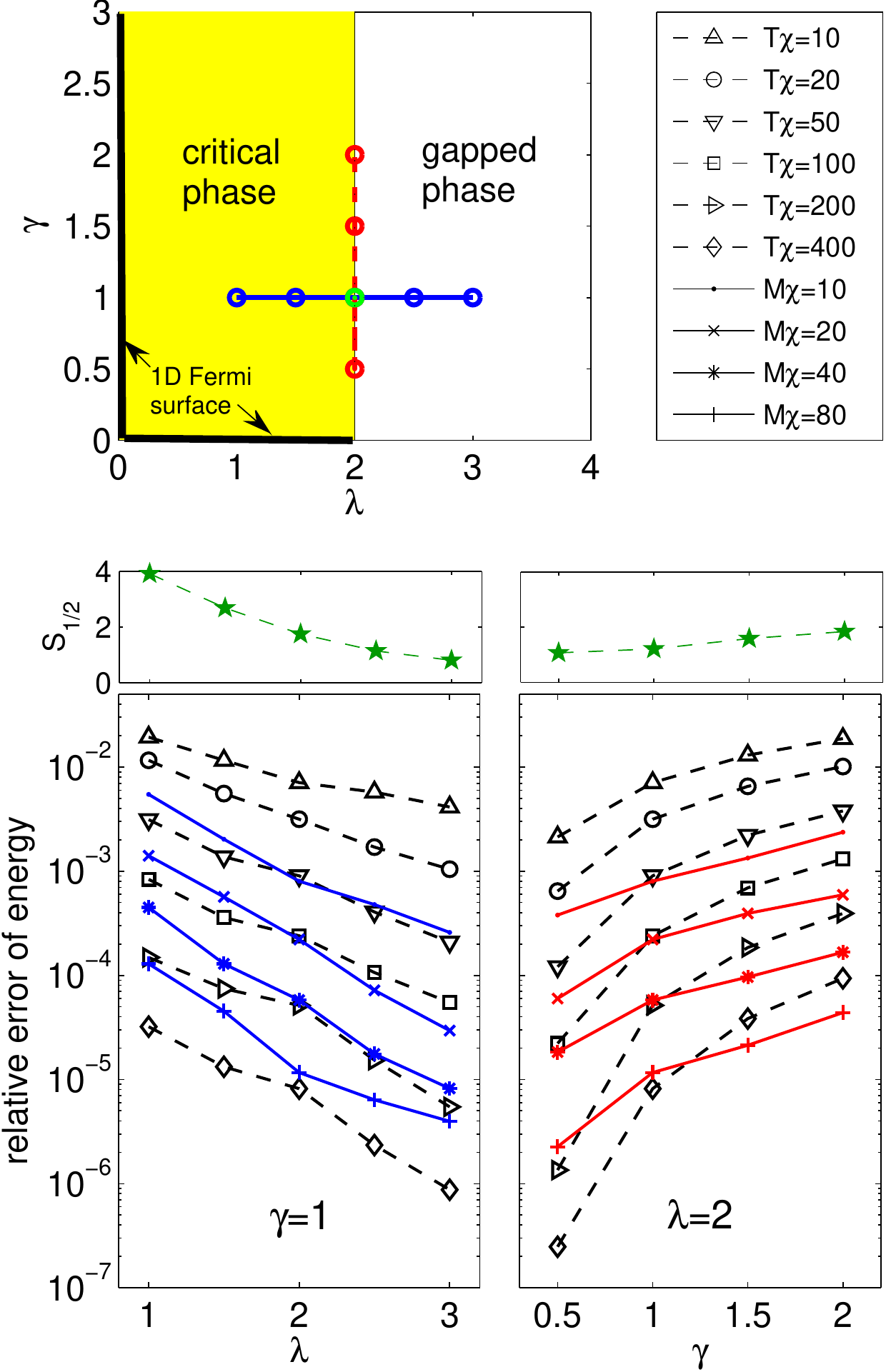}
\caption{(Color online) 
Top left panel: Phase diagram of the free fermion model (\ref{eq:free}). Lower panels: Error in the ground state energy obtained from TTN and MERA simulations of a $6\times 6$ lattice with periodic boundary conditions. The lines correspond to different values of the refinement parameter $\chi$, as indicated in the legend in the top right panel. The accuracy of the simulation results depends on the amount of entanglement in the system, 
which is measured here by the entanglement entropy of half the  system, $S_{1/2}$, plotted in the middle panels.} 
\label{fig:spinless}
\end{center}
\end{figure}

\begin{figure}[htb]
\begin{center}
\includegraphics[width=8cm]{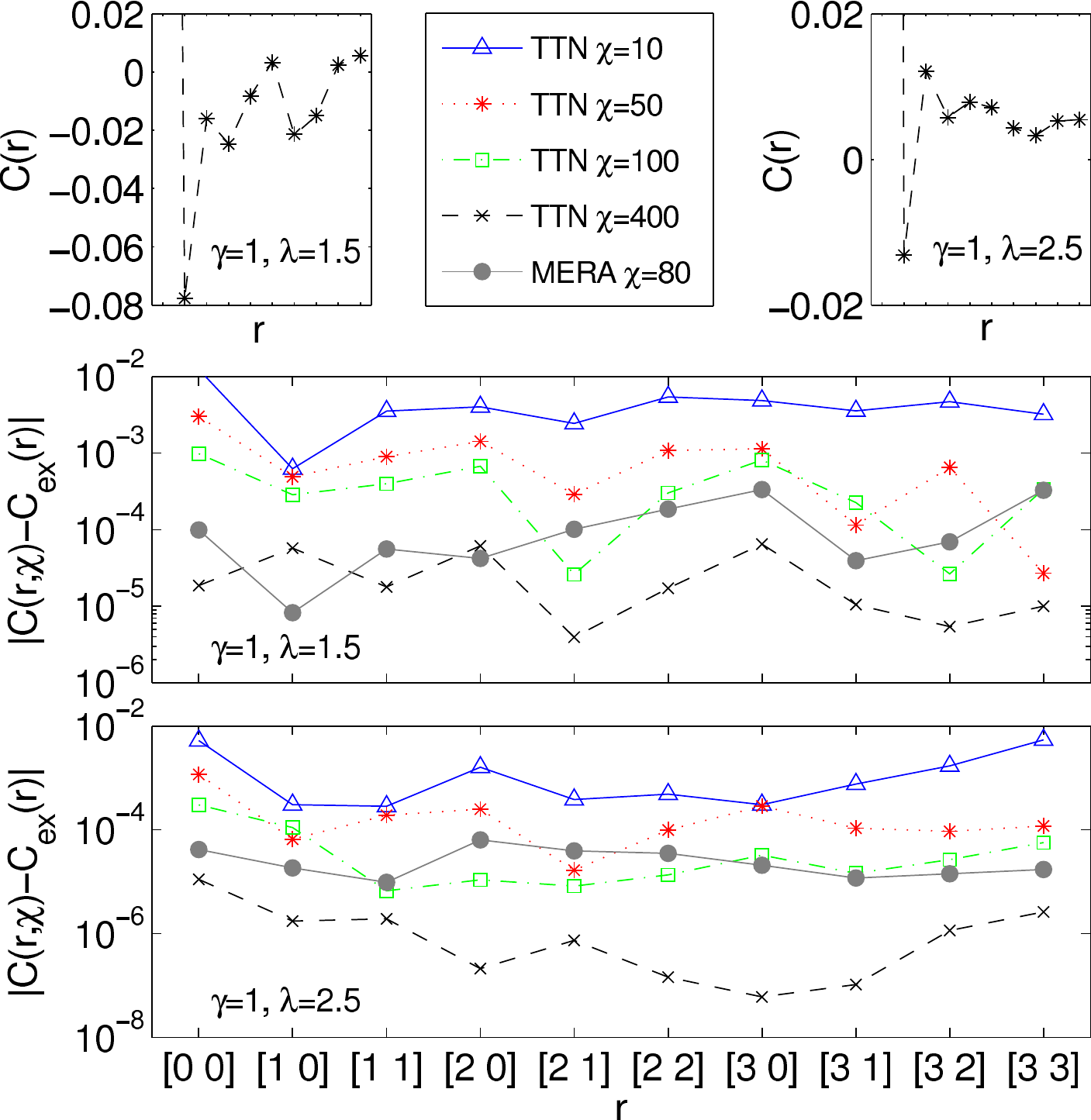}
\caption{(Color online) 
Top panels: Correlation function $C({\vec r})= \langle c_{\vec r_0}^{\dagger} c_{\vec r_0 + \vec r} \rangle$ for $\gamma=1,\lambda=1.5$ (left) and for  $\gamma=1,\lambda=2.5$ (right). The positions $\vec r$ are the same as indicated in the bottom plot. Lower panels: The difference between the simulation result and the exact analytical solution for different values of the refinement parameter $\chi$.  } 
\label{fig:corr}
\end{center}
\end{figure}

We first consider an exactly solvable model of non-interacting spinless fermions in two dimensions given by the Hamiltonian
\begin{equation}
	H_{\mbox{\tiny{free}}} = \sum_{\langle rs \rangle} [c_r^{\dagger}c_s + c_s^{\dagger}c_r - \gamma(c_{r}^{\dagger}c_{s}^{\dagger} + c_{s}c_{r})] - 2\lambda \sum_r c_{r}^{\dagger} c_r,
\label{eq:free}
\end{equation}
with $\lambda$ the chemical potential and $\gamma$ the pairing potential. The phase diagram of this model (see Fig. \ref{fig:spinless}) exhibits a critical (p-wave) superconducting phase for $\gamma>0$, $0 < \lambda < 2$ with two gapless modes, and a gapped superconducting phase for $\gamma>0$, $\lambda>2$. \cite{Li06, sccomment} For $\gamma=0$ the model corresponds to a free fermion system, i.e. a metal (with a one dimensional Fermi surface) for $0<\lambda<2$ and a band-insulator for $\lambda>2$. A metallic phase is also found for $\gamma>0$ and $\lambda=0$.
The lower panels in Fig. \ref{fig:spinless} present the error in the ground state energy of a $6 \times 6$ system as a function of $\gamma$ and $\lambda$, for increasing values of $\chi$. Both TTN and MERA reproduce several significant digits of the exact solution. The middle panels show the entanglement entropy of half the  system,
\begin{equation}
S_{1/2}=- \sum_k \lambda_k \log_2 \lambda_k,
\end{equation}
with $\lambda_k$ the eigenvalues of the reduced density matrix of half the system. The accuracy of the energy is clearly correlated with the amount of entanglement in the system, i.e. the accuracy decreases with increasing $S_{1/2}$. 
Accurate results are also obtained for correlators, $C({\bf r})= \langle c_{\bf r_0}^{\dagger} c_{\bf r_0 + \bf r} \rangle$, as shown in Fig. \ref{fig:corr}.  
Note that the $6 \times 6$ system corresponds to a MERA with only one single layer of isometries and disentanglers, which has a computational cost that scales as $O(\chi^4)$ (times a factor depending on the local dimension $d$ in the lattice ${\cal L}_0$). 
This allows us to use a larger value of $\chi$ than in the MERA with several layers (large systems, see below).

\begin{figure}[htb]
\begin{center}
\includegraphics[width=8cm]{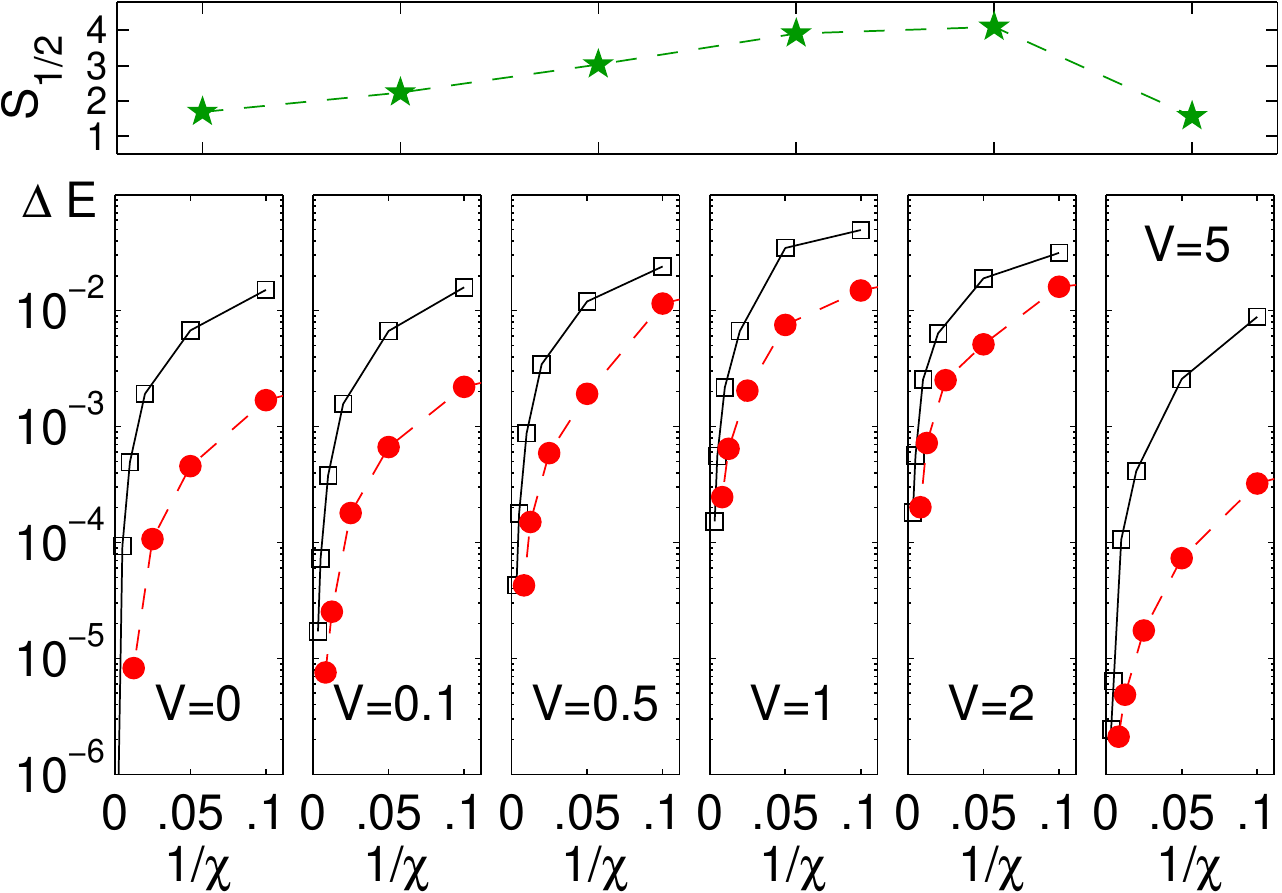}
\caption{(Color online) 
Convergence of the ground state energy of interacting spinless fermions on a $6 \times 6$ lattice with periodic boundary conditions and $\gamma=1, \lambda=2$ for different interaction strengths $V$. The plot shows $\Delta E = E_{\chi}-E_{min}$, the difference between the energy as a function of $\chi$ and the best (lowest) energy obtained by the simulations. Open squares are obtained by the TTN, filled circles by the MERA.} 
\label{fig:ediff}
\end{center}
\end{figure}

\begin{figure}[htb]
\begin{center}
\includegraphics[width=8cm]{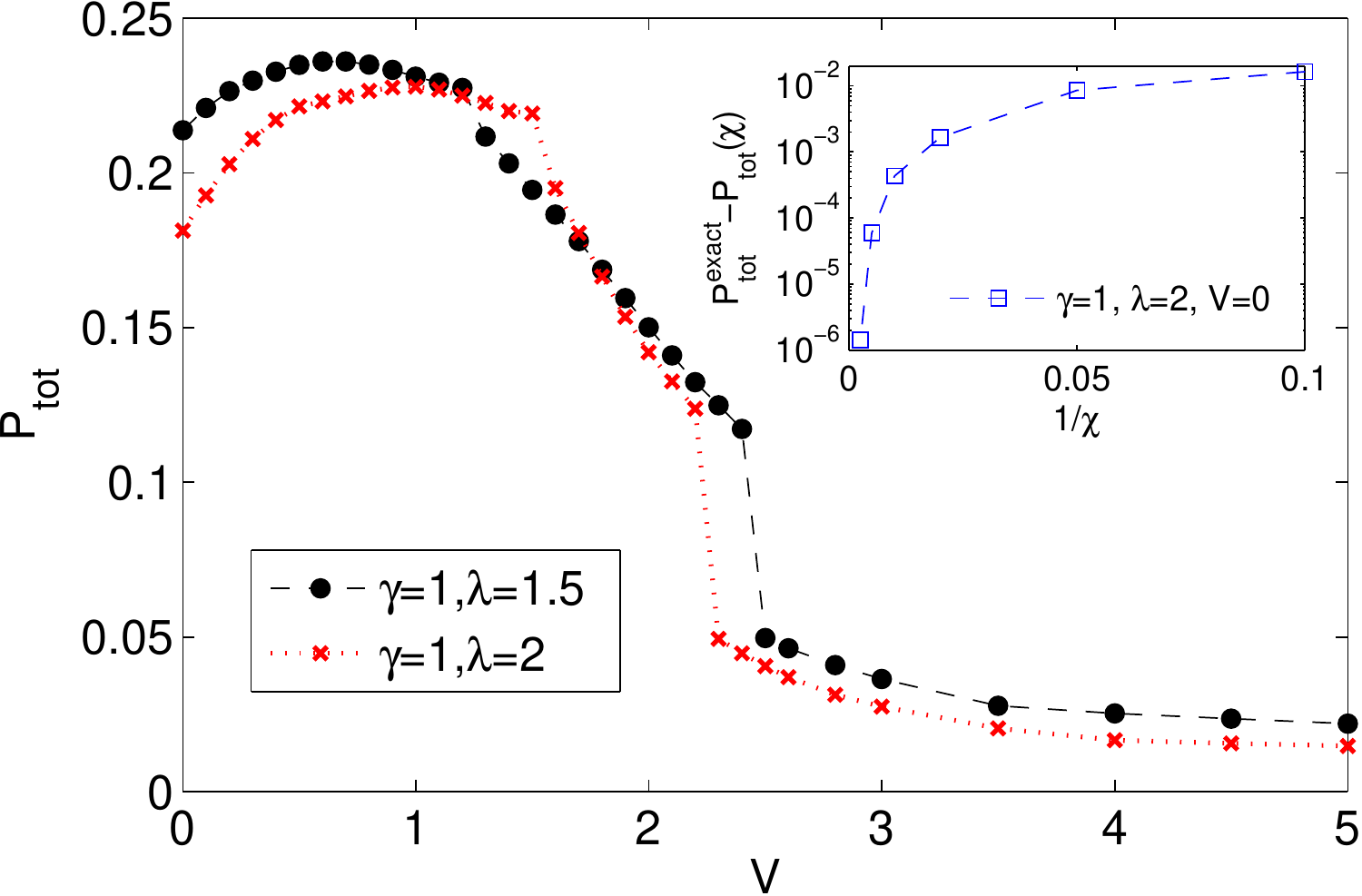}
\caption{(Color online) 
Total pairing amplitude $P_{tot}$ as a function of the interaction strength $V$ of the spinless fermion model \eqref{eq:Hint}, obtained by a TTN with $\chi=200-300$. Larger values of $\chi$ produce corrections between $10^{-3}$ (for $V\approx 1.5$) and $10^{-4}$. The inset shows the convergence of $P_{tot}$ with $\chi$ in the exactly solvable case, $V=0$. } 
\label{fig:scop}
\end{center}
\end{figure}

Next we consider the same Hamiltonian \eqref{eq:free} with an additional nearest-neighbor interaction, 
\begin{equation}
	H_{\mbox{\tiny{int}}} = H_{\mbox{\tiny{free}}} + V \sum_{\langle rs \rangle}  c_r^{\dagger}c_r  c_s^{\dagger}c_s,
\label{eq:Hint}
\end{equation}
which can no longer be solved analytically. We emphasize that the algorithm does not require any particular modification in order to deal with the interaction, since an arbitrary 2-body Hamiltonian can be used as an input to the simulation.
The lower panels in Fig. \ref{fig:ediff} show the convergence of the energy with $\chi$ for different interaction strengths $V$. For small ($V<<1$)  and large ($V>>1$) interaction we find a similar convergence behavior as in the non-interacting case. In both cases $S_{1/2}$ is relatively small, as shown in the upper panel of Fig. \ref{fig:ediff}.
For an interaction strength of the order of the hopping amplitude, $V \sim t \equiv 1$, the convergence with $\chi$ is slower but $\approx 4$ digits of accuracy are still achieved for large $\chi$. Accordingly, the amount of entanglement in the system (measured by $S_{1/2}$) is large in this parameter region.
As another example of a correlation function we computed the pairing amplitude 
\begin{equation}
P({\bf k})= \langle c^\dagger_{\bf k} c^\dagger_{-{\bf k}} \rangle, \quad c^\dagger_{\bf k}= \frac{1}{\sqrt{N}} \sum_{\bf r} c^\dagger_{\bf r} \exp(i {\bf k} {\bf r}). 
\end{equation}
Figure \ref{fig:scop} shows the total pairing amplitude $P_{tot}=\sum_{\bf k} | P({\bf k})|$ as a function of $V$, for two sets of parameters for $\lambda$ and $\gamma$.
Also this quantity converges to the exact solution with increasing $\chi$ in the exactly solvable case, $V=0$, as shown in the inset. A similar convergence behavior is observed in the interacting case (not plotted). A small interaction amplifies the total pairing amplitude, whereas a large interaction tends to suppress the pairing. The sudden jump of both curves around $V \approx 2.2$ could indicate a first order phase transition. A weaker feature is found for $V \approx 1.5$ (crosses) and $V \approx 1.2$ (dots). 

\begin{figure}[htb]
\begin{center}
\includegraphics[width=8cm]{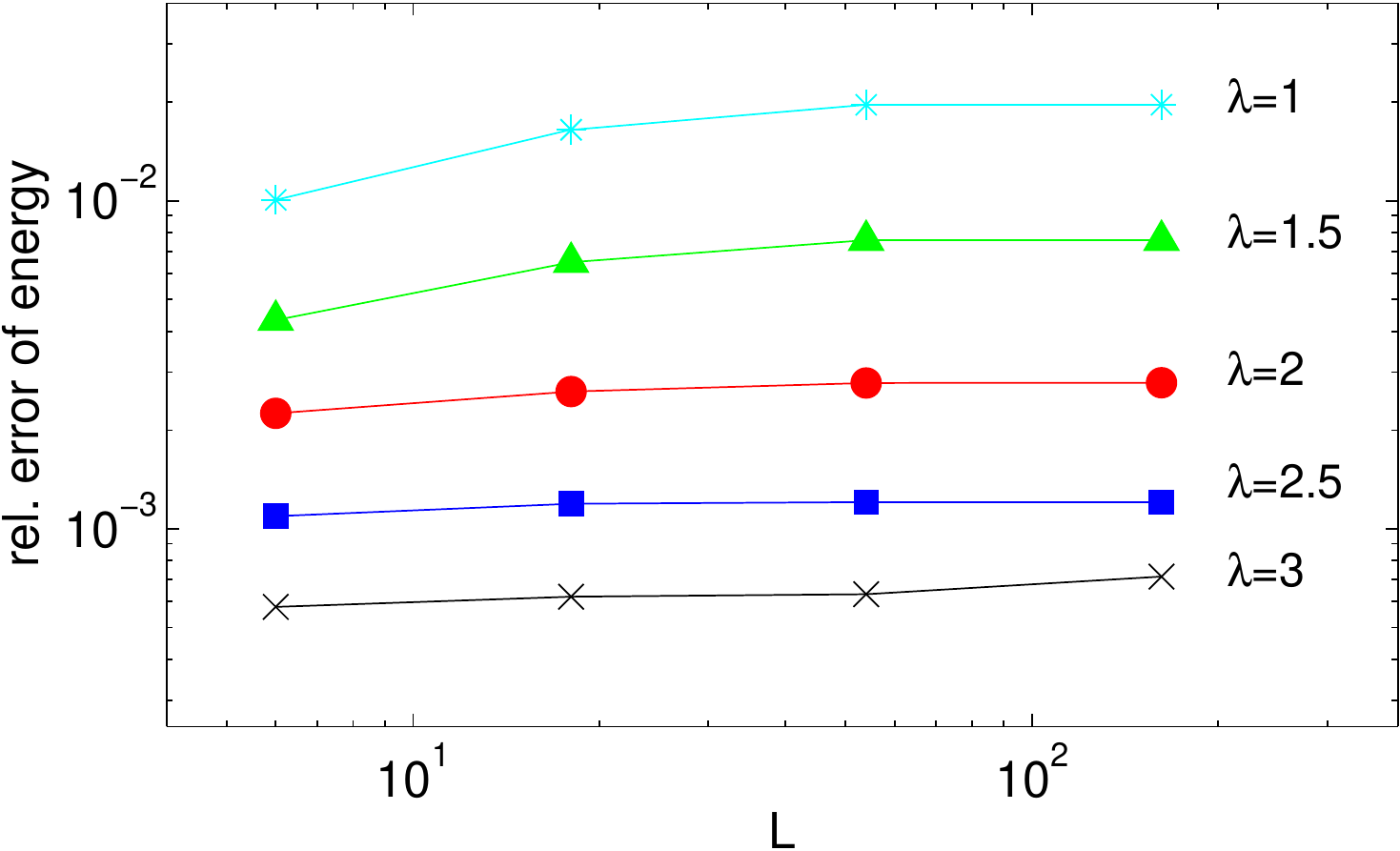}
\caption{(Color online) 
The relative error of the energy as a function of system size obtained by the 2D MERA is of the same order of magnitude for small systems as for large systems, even in the critical regime $\lambda<2$. The simulations are done for $\chi=4$ up to a system size $162 \times 162$, and $V=0$.} 
\label{fig:Ediff}
\end{center}
\end{figure}

Finally, we show that the MERA is scalable in two dimensions. Figure \ref{fig:Ediff} shows the relative error of the energy as a function of system size up to $162\times 162$ for the non-interacting case $V=0$.\cite{comment:scalablemera} For a fixed $\chi=4$ the relative error is of the same order of magnitude for small systems as for large systems, even in the critical regime $\lambda\le 2$. The system size can easily be increased by adding more layers of the MERA, with a cost that only grows logarithmically with the system size for a translational invariant system. \cite{algorithm} 

\section{Conclusion}
\label{sec:conclusion}
To summarize, we have shown that fermionic systems can be addressed in a very similar way as bosonic systems within the formalism of entanglement renormalization. 
We explained how to modify the bosonic MERA in order to deal with fermionic degrees of freedom. To do this we incorporate a $\ZZ$ symmetry into the MERA by using parity preserving tensors, and introduce a fermionic swap gate to account for the exchange of fermionic degrees of freedom, whenever two lines in the tensor network cross.  
 We showed that a fermonic tensor network can be transformed in such a way that the fermionic swap gates do not increase the complexity of a contraction. Thus, an important result is that the complexity of the fermionic MERA is the same as for the bosonic MERA. 
The present formalism to deal with fermionic systems was originally developed specifically for the TTN and the MERA in mind but, in its present formulation, can be applied to arbitrary tensor networks. The steps to be followed are surprisingly simple: given a tensor network ansatz, such as PEPS, one must choose $\ZZ$ symmetric (i.e. parity invariant) tensors and, when contracting the tensor network, replace crossings with fermionic swap gates. This procedure is exemplified in Ref. \onlinecite{FCTM} for infinite PEPS.
 
Here we have presented benchmark results for the 2D MERA and the TTN for spinless fermions, both non-interacting (exactly solvable) and interacting. We have also shown that the 2D MERA  is scalable by simulating lattices made of up to 162x162 sites. 
      
In general, the efficiency of the MERA depends on the amount of entanglement in the ground state of the system. Accordingly, as discussed in Ref. \onlinecite{FreeFermions} for free fermions, gapped systems appear typically as the easiest to simulate. They are followed by critical phases with a finite number of zero modes (e.g. Dirac modes), which are more entangled but still follow an area law for the entanglement entropy,\cite{Wolf06, Gioev06, Li06, Barthel06} which a MERA with the same $\chi$ at each level of coarse-graining can reproduce.\cite{FreeFermions} The most challenging systems are metals with a one dimensional Fermi surface, i.e. an infinite number of zero modes. These are the most entangled systems, with a multiplicative logarithmic correction to the area law.\cite{Wolf06, Gioev06, Li06, Barthel06}

The efficiency of the algorithm can be substantially improved by making use of symmetries (e.g. $SU(2)$, $U(1)$, etc.) of a model, \cite{Singh09} and by variational Monte Carlo sampling techniques. \cite{Schuch07, Sandvik07} This is important in order to increase the maximal affordable $\chi$, which for large systems is still small at present.
A higher accuracy can also be achieved by choosing an optimal structure of the 2D MERA depending on the problem considered. An example of an improved coarse-graining scheme was presented in Ref. \onlinecite{2D}.

We believe that the fermionic MERA will help to shed new light into long standing questions in strongly correlated fermion systems. Work in progress includes the study of the ground state phase diagram of the tJ and the Hubbard model, and generalizations to anyonic systems.\cite{Aguado09}

We thank R. Pfeifer and L. Tagliacozzo for useful discussions, and S. Haas, L. Ding and N. Ali for clarifications concerning the free fermion model \eqref{eq:free}. 
Support from the Australian Research Council (APA, FF0668731, DP0878830) is acknowledged.

NOTE: Short after Ref. \onlinecite{Corboz09} (of which the present paper is an extended version) was made available online, a largely equivalent approach has been independently presented by C. Pineda, T. Barthel and J. Eisert in Ref. \onlinecite{Pineda09}.

\appendix

\section{The fermionic swap gate}
\label{app:swap}
In this appendix we show that the fermionic swap gate implements the anticommutation of fermionic operators, and make connection to the Jordan-Wigner transformation used in Ref. \onlinecite{Corboz09}. 

\begin{figure}[!htb]
\begin{center}
\includegraphics[width=7cm]{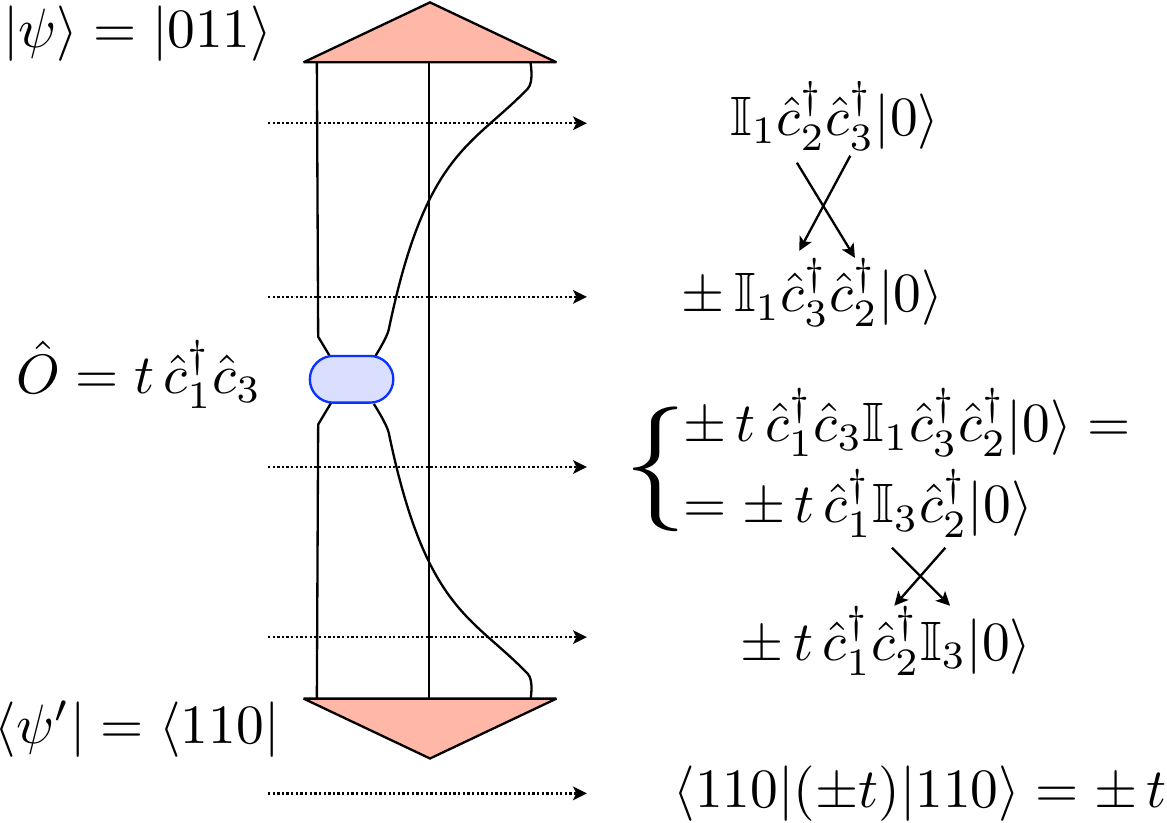}
\caption{(Color online) 
A diagram representing the matrix element $\expect{\psi' | \hO | \psi}$. Reading the diagram from top to bottom, we obtain a prescription of how to calculate the matrix element. Crossing lines imply that the operators carried by the line are exchanged, which results in a negative sign in the fermionic case if two creation (or annihilation) operators are exchanged. The resulting matrix element is $+t$ in the bosonic and $-t$ in the fermionic case.  } 
\label{fig:diag3}
\end{center}
\end{figure}

In a fermionic MERA the wires in the quantum circuit carry fermionic degrees of freedom, and all gates and operators can be expanded as a product of fermionic creation and annihilation operators. These operators anticommute instead of commuting as in the bosonic case.
To illustrate this essential difference between a fermionic and a bosonic MERA (for hardcore bosons) we consider the computation of a matrix element of an operator $\hO$ acting on sites 1 and 3 in a three-site system, shown in Fig. \ref{fig:diag3}. The full Hilbert space is spanned by the basis states 
\begin{equation}
\label{eq:3sb}
\ket{i_1 i_2 i_3} \equiv   \hc_1^{\dagger i_1} \hc_2^{\dagger i_2} \hc_3^{\dagger i_3} \ket{0}
\end{equation}
where $i_k \in \{0,1\}$ and $\hc_k^{\dagger}$ creates a particle on site $k$, obeying the commutation relations
\begin{equation}
\label{eq:anticomm}
\left[\hc_i,\hc_j \right]_{\pm} =0, \quad \left[\hc_i,\hcdag_j\right]_\pm = \delta_{ij}, \quad i,j \in \{1,2,3\}.
\end{equation}
In the bosonic case $(+)$ operators on different sites commute, whereas in the fermionic case $(-)$ they anticommute. 
Let us consider the example of a hopping operator $\hO=t \hcdag_1 \hc_3$, and the states $\ket{\psi}=\ket{0 1 1}$ and $\bra{\psi'}=\bra{1 1 0}$. Figure \ref{fig:diag3} provides a graphical prescription of how to compute the matrix element, 
\begin{equation}
 \expect{\psi' | \hO | \psi} = \expect{1 1 0 | t \hcdag_1 \hc_3 | 0 1 1} =  t \expect{0 | \hc_2 \hc_1 \hcdag_1 \hc_3 \hcdag_2 \hcdag_3 | 0}. 
\end{equation}
For a bosonic system this matrix element is simply $t$, because operators on different sites commute. However, in the fermionic case a minus sign appears from exchanging $\hcdag_2$ with $\hcdag_3$, as shown in Fig. \ref{fig:diag3}. Thus, a line crossing is a graphical representation of the exchange of operators, and a negative sign results if both lines carry an odd number of fermionic creation/annihilation operators. 

\begin{figure}[!htb]
\begin{center}
\includegraphics[width=7cm]{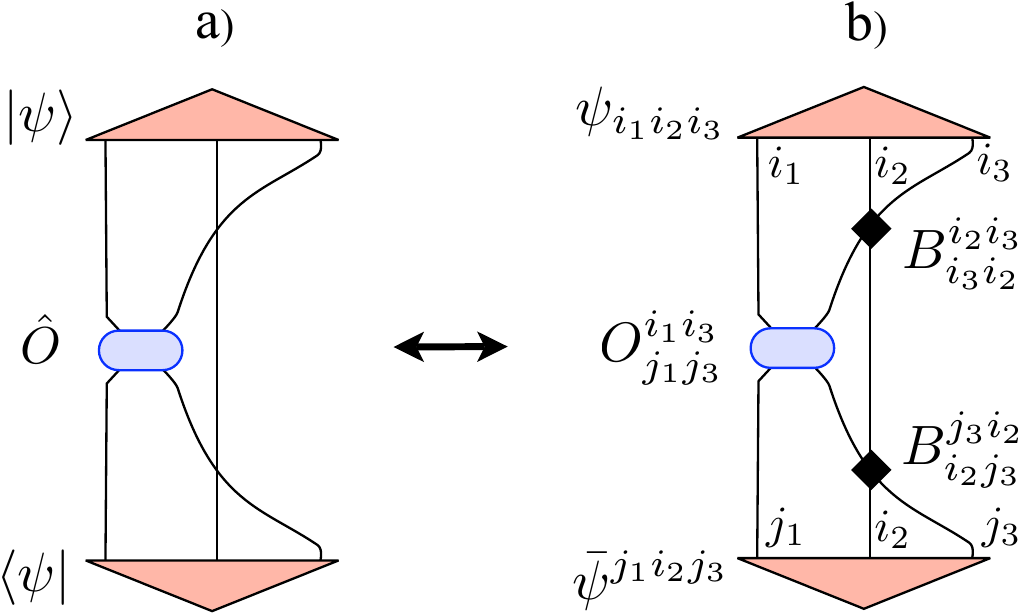}
\caption{(Color online) 
a) A general expectation value of a two-body operator acting on sites 1 and 3. As usual, the crossing of lines implies exchanging the operators carried by the lines. b) Mapping to a tensor network by replacing each line crossing by a swap gate, which implements the anticommutation of fermionic operators.} 
\label{fig:jw}
\end{center}
\end{figure}
More generally, consider an arbitrary (parity preserving operator) $\hO$ acting on sites 1 and 3 and a state $\ket{\psi}$ expanded in the local basis,
\begin{eqnarray}
\label{eq:expansion}
\ket{\psi}=\sum_{i_1 i_2 i_3} \psi_{i_1 i_2 i_3} \ket{i_1 i_2 i_3}, \quad \hat{O}= \! \! \sum_{i_1 i_3 j_1 j_3} \! O^{i_1 i_3}_{j_1 j_3} \hc_{1}^{\dagger i_1} \hc_{3}^{\dagger i_3} \hc_{3}^{j_3} \hc_1^{j_1} \nonumber.
\end{eqnarray}
The expectation value is given by
\begin{eqnarray}
\label{eq:E2}
 \expect{\psi|\hat O | \psi} &=&\sum_{i_1 i_2 i_3  \atop j_1 j_3 } \bar{\psi}^{j_1 i_2 j_3} O^{i_1 i_3}_{j_1 j_3} \psi_{i_1 i_2 i_3}  \\
 &&\times  \expect{0| (\hc_{3}^{j_3}  \hc_{2}^{i_2}  \hc_{1}^{j_1} ) ( \hc_{1}^{\dagger j_1}  \hc_{3}^{\dagger j_3}  \hc_{3}^{i_3}  \hc_{1}^{i_1}  )( \hc_{1}^{\dagger i_1} \hc_{2}^{\dagger i_2} \hc_{3}^{\dagger i_3} )| 0} \nonumber  
\end{eqnarray}
In the bosonic case the second line in Eq. \eqref{eq:E2} is always 1, because the bosonic operators commute, and the expected value is simply obtained by multiplying the tensors in the first line together. In the fermionic case we again have to swap operators as indicated by the crossing lines in Fig. \ref{fig:jw}a). More conveniently, we replace each crossing by a swap gate introduced in Eq. \eqref{eq:swapgate}, which accounts for the anticommutation rules of the fermionic operators, so that the expectation value becomes
\begin{equation}
\expect{\psi|\hat O | \psi} =\sum_{i_1 i_2 i_3  \atop j_1 j_3 } \bar{\psi}^{j_1 i_2 j_3} B^{j_3 i_2}_{i_2 j_3} O^{i_1 i_3}_{j_1 j_3} B^{i_2 i_3}_{i_3 i_2} \psi_{i_1 i_2 i_3}.
\end{equation}
Therefore, replacing each line crossing by a swap gate transforms the diagram in Fig. \ref{fig:jw}a) into the tensor network shown in Fig. \ref{fig:jw}b), which we can contract  as explained in Sec. \ref{sec:objects}.

\begin{figure}[!htb]
\begin{center}
\includegraphics[width=8cm]{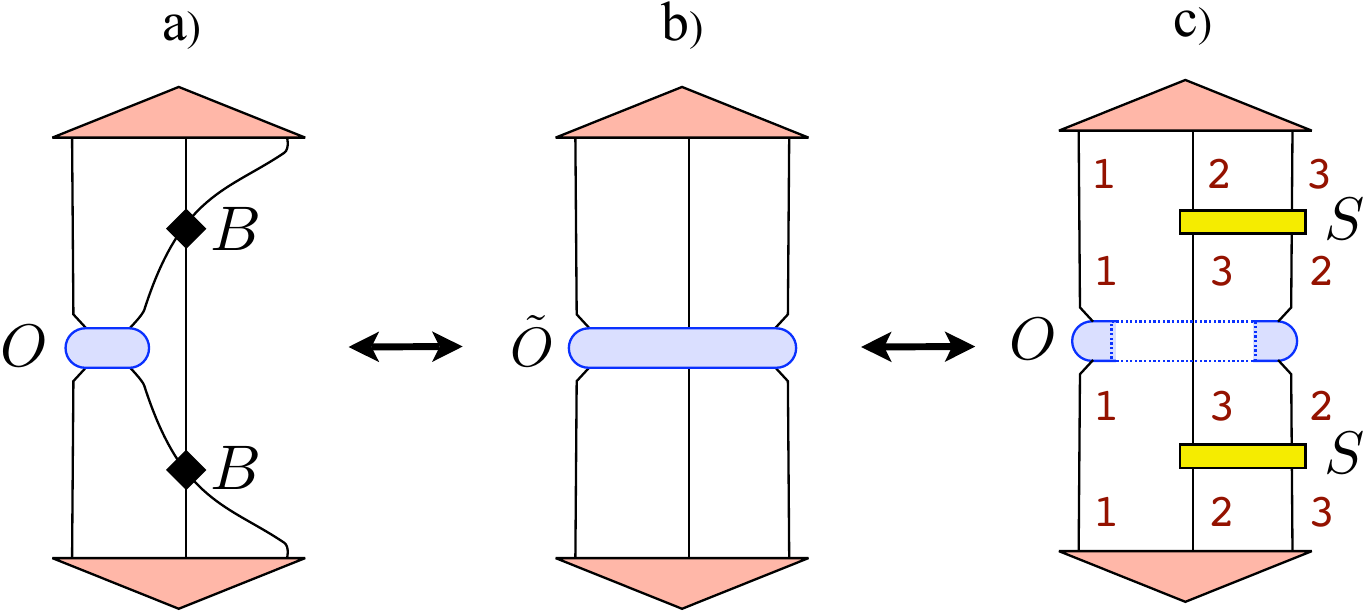}
\caption{(Color online) 
Three equivalent approaches to compute an expectation value of a fermionic two-body operator acting on sites 1 and 3: 
a) by using swap gates as presented in this work, b) by using a Jordan-Wigner transformation of the fermionic operator $O$ into a three-site spin (bosonic) operator $\tilde O$, c) by changing the Jordan-Wigner order of sites 2 and 3 by a gate $S$ (cf. text), so that the operator $O$ acts on contiguous sites (with respect to the Jordan-Wigner order as indicated by the numbers). } 
\label{fig:approaches}
\end{center}
\end{figure}

If we incorporate the swap gates into the operator $\hO$ we end up with the three-site operator shown in Fig. \ref{fig:jw}b), which is nothing but the Jordan-Wigner transformation of operator $\hO$, i.e. the fermionic operator mapped to (bosonic) spin variables. This approach was used to introduce the fermionic MERA in Ref. \onlinecite{Corboz09}, but it is important to point out, that it is \textit{equivalent} to the fermionic MERA presented in this paper. A third equivalent approach (but yet another point of view) is to change the Jordan-Wigner order of the lattice sites such that operator $\hO$ acts on contiguous sites, as illustrated in Fig. \ref{fig:approaches}. The gate $S$ to change the Jordan-Wigner order corresponds to the swap gate $B$  introduced in Eq. \eqref{eq:swapgate}, 
except that the lines do not cross, i.e. $S^{i_1 i_2}_{j_1 j_2}=B^{i_1 i_2}_{j_2 j_1}$. We summarize the three equivalent approaches in appendix \ref{sec:ea}.

\section{Proof of the "jump" move}
\label{app:jump}
The "jump" move introduced in Sec. \ref{sec:fswap} allows us to drag a line over a tensor, as for example shown in Fig. \ref{fig:jump}. The validity of this rule originates from the particular form of the swap gate and from the fact that all tensors preserve parity, as we explain in the following.

\begin{figure}[htb]
\begin{center}
\includegraphics[width=8.5cm]{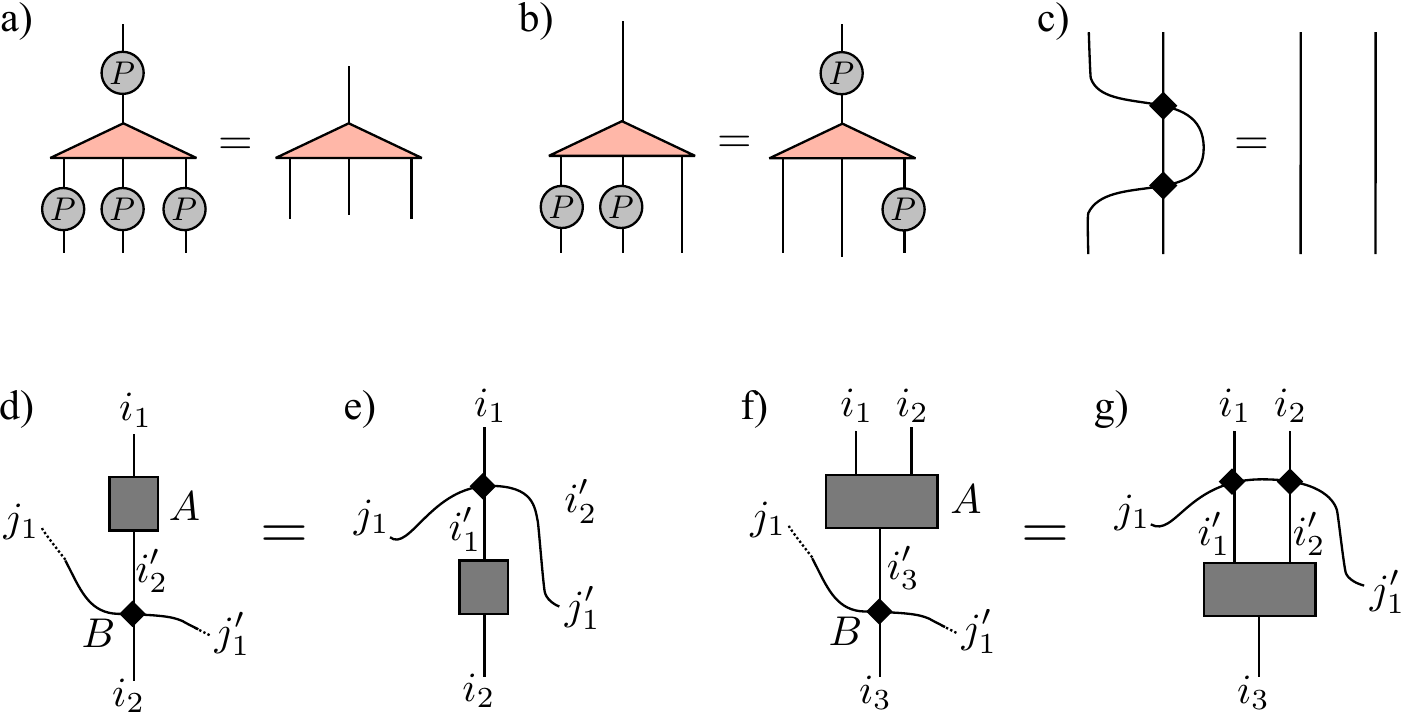}
\caption{(Color online) 
a) A parity preserving tensor remains invariant when applying the parity operator $P$ on each leg. b) This equality follows from $P^2=1$, i.e. the parity operator squared is the identity. 
c) Two subsequent line crossings is equivalent to the identity. d) and e) "Jump" move over a tensor with two legs, corresponding to Eq. \eqref{eq:jump1}. f) and g) "Jump" move over a tensor with three legs, explained in Eq. \eqref{eq:jump2}. } 
\label{fig:jumpproof}
\end{center}
\end{figure}

First note that, by definition, a parity preserving tensor remains invariant when the parity operator $P$ acts on all of its legs, as illustrated in Fig. \ref{fig:jumpproof}a). It then follows from $P^2=1$ that acting with $P$ on a set of legs of a tensor is equivalent to acting with $P$ on the complementary set of legs of the tensor, see Fig. \ref{fig:jumpproof}b). 
Next we consider a two-legged tensor $A$ and a swap gate $B$ (see Eq.  \eqref{eq:swapgate}) shown in Fig. \ref{fig:jumpproof}d). Contracting $A$ and $B$ amounts to
\begin{eqnarray}
\label{eq:jump1}
T^{j_1 i_1}_{i_2 j_1'}= \sum_{i_2'} A^{i_1}_{i_2'} B^{j_1 i_2'}_{i_2 j_1'} = \delta_{j_1,j_1'}  A^{i_1}_{i_2} S(P(j_1),P(i_2)) \nonumber \\
= \delta_{j_1,j_1'} A^{i_1}_{i_2}  S(P(j_1),P(i_1))=  \sum_{i_1'}  B^{j_1 i_1}_{i_1' j_1'} A^{i_1'}_{i_2}
\end{eqnarray}
where we made use of the parity symmetry of $A$, i.e. that acting with the parity operator $P$ on the lower leg of $A$ is equivalent to acting with $P$ on the upper leg. The final expression in Eq. \eqref{eq:jump1} is nothing but the swap gate now acting on the upper leg of tensor $A$ as shown in Fig. \ref{fig:jumpproof}e).
In a similar way one can proof the "jump" move for a three-legged tensor shown in Fig. \ref{fig:jumpproof}f) and g) by contracting the tensor networks and using 
\begin{eqnarray}
\label{eq:jump2}
S(P(j_1),P(i_3))&=&S(P(j_1),P(i_1)P(i_2)) \\
&=&S(P(j_1),P(i_1))S(P(j_1),P(i_2)). \nonumber
\end{eqnarray}
The first equality again follows from the parity symmetry, and the second equality can be easily verified.
This property can be extended to tensors with more than 3 legs by applying the last identity recursively, i.e.
\begin{equation}
S(P(j_1), \prod_k P(i_k))= \prod_k S(P(j_1), P(i_k))
\end{equation}
Finally, we show that two subsequent line crossings of the same lines is simply the identity, as shown in Fig. \ref{fig:jumpproof}c). 
This follows from multiplying two swap gates together,
\begin{eqnarray}
T^{i_1 i_2}_{k
_1 k_2}&=& \sum_{j_1 j_2} B^{i_1 i_2}_{j_2 j_1} B^{j_2 j_1}_{k_1 k_2} = \\ \nonumber
&=& \delta_{i_1,k_1}\delta_{i_2,k_2} \left(S(P(i_1),P(i_2)\right)^2\\ \nonumber 
&=& \delta_{i_1,k_1}\delta_{i_2,k_2}.
\end{eqnarray}
By combining above rules an arbitrary jump move can be performed. 

\section{Causal cone of the fermionic MERA}

\label{app:cc}
It is important to notice that the fermionic MERA has the same causal cone as the bosonic MERA (cf. Fig. \ref{fig:causalcone}). This can be understood as a consequence of the "jump" move, as shown in Fig. \ref{fig:jump}. An arbitrary diagram can be modified in such a way that no line crosses the outgoing wires of a gate that lies outside the causal cone. The gate with its conjugate can then be replaced by the identity thanks to Eqs. \eqref{eq:u} and \eqref{eq:w}, as done for example in the diagram in Fig. \ref{fig:jump} with the isometry in the middle.

Another way to arrive to this conclusion is to consider the MERA where we expand each gate and the Hamiltonian in its local fermionic basis. An expectation value of an operator $\hO$ is of the form
\begin{eqnarray}
\label{eq:cc}
\expect{\hO}\!=\! \expect{0 | \hwdag_{N_w} \! \! \dots \! \hwdag_1 \hudag_{N_u} \!\! \dots \! \hudag_1 \hO \hu_1 \! \dots \! \hu_{N_u} \hw_1 \! \dots \! \hw_{N_w} | 0}
\end{eqnarray}
where we consider only one layer of the MERA for simplicity, and $N_w$ and $N_u$ are the number of isometries and disentanglers in the layer, respectively. Because each gate is parity preserving, its expansion consists only of terms with an even number of creation/annihilation operators. As a consequence two parity preserving gates with disjoint supports (i.e. gates acting on different sites) commute. Thus, to simplify Eq. \ref{eq:cc} we can first commute each disentangler $\hu_k$ that lies outside the causal cone of operator $\hO$ to the left to annihilate with its conjugate, and then proceed similarly with the isometries $\hw_k$, so that only gates inside the causal cone of operator $\hO$ remain. For example, if $\hu_1$ lies outside the causal cone, $[ \hO, \hu_1]=0$ and $\hudag_1 \hu_1 = 1$ annihilate. This generalizes straightforwardly to several layers of isometries and disentanglers as in Fig. \ref{fig:causalcone}. \\

\section{Equivalent approaches}
\label{sec:ea}
In this appendix we briefly review the formulation of the fermionic MERA originally presented in Ref. \onlinecite{Corboz09}, together with an alternative formulation also outlined in Ref. \onlinecite{Corboz09}, and compare it to the simplified formulation presented in this paper.

\textit{1) Fixed Jordan-Wigner order:} The formalism introduced in Ref. \onlinecite{Corboz09} is based on the Jordan-Wigner transformation (with a fixed Jordan-Wigner order), which maps all fermionic operators into spin operators with string of Z's (the $\sigma_z$ Pauli matrix in case of a two dimensional local Hilbert space). An example of such an operator is shown in Fig. \ref{fig:approaches}b). These strings of Z's are coarse-grained locally by using fermionic disentanglers and isometries. The fermionic trace allows us to dispense with the string of Z's, as noticed in Ref. \onlinecite{Corboz09}. For example, the reduced density matrix of sites $1$ and $3$ of a system with three sites is computed as
%
\begin{eqnarray}
 \rho_{1 3} &=& \mbox{ftr}_{2} (\rho_{1 2 3 }) \nonumber \\
 &\equiv& \sum_{\alpha=0,1} {\cal P}_{1'}^{(\alpha)} \tr_{2} (\rho_{1 2 3 }) {\cal P}_{1 }^{(\alpha)} \nonumber \\ 
 &+& \sum_{\alpha=0,1} {\cal P}_{1}^{(\alpha)} \tr_{2} (\rho_{1 2 3}Z_{2}) {\cal P}_{1}^{(1-\alpha)},
\end{eqnarray}
where ${\cal P}_{r}^{(0)}$ and ${\cal P}_{r}^{(1)}$ project onto the even and odd parity sectors of site $r$, and $\rho_{1 2 3}$ is the density matrix of the full system. Indeed, one can check that
\begin{eqnarray}
\tr(A_1Z_2 B_3 \rho_{123}) &=& \tr (A_1B_3 \rho_{13}) \\
\tr(C_1 I_2 D_3 \rho_{123}) &=& \tr (C_1 D_3 \rho_{13})
\end{eqnarray}
for $A$, $B$ parity changing operators (odd parity operators) and $C$, $D$ parity preserving operators (even parity operators)

\textit{2) Changing the Jordan-Wigner order:} The use of a fermionic trace amounts to effectively changing the Jordan-Wigner order in such a way that, in the new order, the sites to be kept after tracing out are contiguous sites. Similarly, before applying a specific operator, the Jordan-Wigner order is changed accordingly so that the operator acts on contiguous sites in the new order, as shown, for instance, in Fig. \ref{fig:approaches}c). The same holds for isometries and disentanglers. This was pointed out in Ref. \onlinecite{Corboz09} (see also Refs. \onlinecite{Pineda09, Barthel09}).

\textit{3) Crossing of lines carrying fermionic degrees of freedom:} As discussed in appendix \ref{app:swap}, the above approaches are also equivalent to the one presented in this paper. The latter has the advantage of completely dispensing with the Jordan-Wigner order, and its implementation on generic tensor network algorithms for 1D lattices with periodic boundary conditions (e.g. MPS, TTN, MERA), and 2D lattices (e.g. PEPS, TTN, MERA) is straightforward.


\begin{thebibliography}{99}
\bibitem{comment:d} Note that the local dimension $d$ of the original lattice, in which the Hamiltonian is defined, is usually smaller than the local dimension $\chi$ of a coarse-grained site.  In this case the computational cost to compute ascending/descending superoperators and environments for the first layer of the MERA is smaller than for the higher layers.
\bibitem{rem:absorbing} Note that in some cases a fermionic swap gate can only be absorbed in a later stage of the contraction, i.e. not necessarily at the beginning.
\bibitem{comment:scalablemera} In this example we study the non-interacting case in order to be able to compare with the exact result. (No exact results are available for the interacting case for large systems). The energy as a function of $\chi$ behaves similarly in the interacting case as in the non-interacting case, but with a slower convergence for intermediate interaction strengths $V \sim 1$ where the amount of entanglement in the system is large (cf. Fig. \ref{fig:ediff}).
\end{thebibliography}
\end{document}